%% file: article_springer.tex
\documentclass[pdflatex,sn-basic]{sn-jnl}

\usepackage{graphicx}%
\usepackage{multirow}%
\usepackage{amsmath,amssymb,amsfonts}%
\usepackage{amsthm}%
\usepackage{mathrsfs}%
\usepackage[title]{appendix}%
\usepackage{xcolor}%
\usepackage{textcomp}%
\usepackage{manyfoot}%
\usepackage{booktabs}%
\usepackage{array}
\usepackage{algorithm}%
\usepackage{algorithmicx}%
\usepackage{algpseudocode}%
\usepackage{listings}%

\theoremstyle{thmstyleone}%
\theoremstyle{thmstyletwo}%

\theoremstyle{thmstylethree}%

\def\M87{M\,87}
\def\SgrA{Sgr\,A*}

\def\3C279{3C\,279}

\def\msun{$M_\odot$}
\usepackage{ulem}
\usepackage{comment}
\usepackage[T1]{fontenc}
\usepackage{longtable}
\makeatletter
\input{aas_macros.sty}
\def\ref@jnl#1{{\jnl@style#1\ }}
\makeatother

\graphicspath{ {./fig/} }
\setcitestyle{aysep={}}

\begin{document}
\title[M\,87: a cosmic laboratory for deciphering BH accretion and jet formation]{M\,87: a cosmic laboratory for deciphering black hole accretion and jet formation}

\author*[1,2,3]{\fnm{Kazuhiro} \sur{Hada}}\email{hada@nsc.nagoya-cu.ac.jp}

\author[4]{\fnm{Keiichi} \sur{Asada}}\email{asada@asiaa.sinica.edu.tw}

\author[5,4]{\fnm{Masanori} \sur{Nakamura}}\email{nakamrms-g@hachinohe.kosen-ac.jp}

\author[6,7]{\fnm{Motoki} \sur{Kino}}
\email{motoki.kino@gmail.com}

\affil*[1]{\orgdiv{Graduate School of Science}, \orgname{Nagoya City University}, \orgaddress{\street{Yamanohata 1, Mizuho-cho, Mizuho-ku}, \city{Nagoya}, \postcode{467-8501}, \state{Aichi}, \country{Japan}}}

\affil*[2]{\orgdiv{Mizusawa VLBI Observatory}, \orgname{National Astronomical Observatory of Japan}, \orgaddress{\street{2-12 Hoshigaoka-cho, Mizusawa}, \city{Oshu}, \postcode{023-0861}, \state{Iwate}, \country{Japan}}}

\affil[3]{\orgdiv{Astronomical Science Program}, \orgname{The Graduate University for Advanced Studies (SOKENDAI)}, \orgaddress{\street{2-21 Osawa}, \city{Mitaka}, \postcode{181-8588}, \state{Tokyo}, \country{Japan}}}

\affil[4]{\orgdiv{Institute of Astronomy and Astrophysics}, \orgname{Academia Sinica}, \orgaddress{\street{11F of Astronomy-Mathematics Building, AS/NTU No. 1, Sec. 4, Roosevelt Road}, \city{Taipei}, \postcode{10617}, \state{Taiwan}}}

\affil[5]{\orgdiv{Department of General Science and Education}, \orgname{National Institute of Technology, Hachinohe  College}, \orgaddress{\street{16-1 Uwanotai, Tamonoki}, \city{Hachinohe}, \postcode{039-1192}, \state{Aomori}, \country{Japan}}}

\affil[6]{\orgname{Kogakuin University of Technology \& Engineering, Academic Support Center}, \orgaddress{\street{2665-1 Nakano-machi}, \city{Hachioji}, \postcode{192-0015}, \state{Tokyo}, \country{Japan}}}

\affil[7]{\orgname{National Astronomical Observatory of Japan}, \orgaddress{\street{2-21 Osawa}, \city{Mitaka}, \postcode{181-8588}, \state{Tokyo}, \country{Japan}}}


\abstract{
Over the past decades, there has been significant progress in our understanding of accreting supermassive black holes (SMBHs) that drive active galactic nuclei (AGNs), both from observational and theoretical perspectives. 
As an iconic target for this area of study, the nearby giant elliptical galaxy M\,87 has received special attention thanks to its proximity, large mass of the central black hole and bright emission across the entire electromagnetic spectrum from radio to very-high-energy $\gamma$-rays. In particular, recent global millimeter-very-long-baseline-interferometer observations towards this nucleus have provided the first-ever opportunity to image the event-horizon-scale structure of an AGN, opening a new era of black hole astrophysics. On large scales, M\,87 exhibits a spectacular jet propagating far beyond the host galaxy, maintaining its narrowly collimated shape over seven orders of magnitude in distance. Elucidating the generation and propagation, as well as the internal structure, of powerful relativistic jets remains a longstanding challenge in radio-loud AGNs. M\,87 offers a privileged opportunity to examine such a jet with unprecedented detail. 

In this review, we provide a comprehensive overview of the observational knowledge accumulated about the M\,87 black hole across various wavelengths. We cover both accretion and ejection processes at spatial scales ranging from outside the Bondi radius down to the event horizon. By compiling these observations and relevant theoretical studies, we aim to highlight our current understanding of accretion and jet physics for this specific object.}




\maketitle 

\pagestyle{myheadings}
\markright{K. Hada et al.} 
{\small
\setcounter{tocdepth}{3}
\tableofcontents
}

\clearpage
\input{table_symbols}

\clearpage
\section{Introduction}\label{sec:intro}
\subsection{General context}\label{ssec:general}

A fraction of galaxies emit an immense amount of energy from the compact central regions, often surpassing the combined brightness of their entire host galaxies. These are known as active galactic nuclei (AGNs), and it is widely accepted that they are powered by the accretion of material onto supermassive black holes (SMBHs) at their cores~\citep{Lyndenbell69, Rees1984}. 
The quest to understand the nature of AGNs and SMBHs has persisted since their initial discovery and remains a paramount focus of the current astrophysics. Furthermore, there is growing evidence that the energetic activities of AGN profoundly influence the formation and cosmological evolution of galaxies and galaxy clusters, as represented by the empirical correlation between the stellar velocity dispersion of galaxy bulges and the mass of central SMBHs~\citep{Magorrian1998, Kormendy2013}. This feedback mechanism may manifest through the intense radiation emitted by luminous accretion disks or the mechanical energy of powerful jets/outflows emanating from the nuclei.

AGNs display diverse observational properties, with various types and subclasses distinguished based on different features. While some of this diversity can be attributed to an orientation effect, where the central engine remains essentially identical, intrinsic differences in the central object also contribute to the range of AGN activity. Most notably, AGNs are divided into two distinct classes based on the ratio of radio to optical luminosity (i.e., radio loudness): radio-quiet (RQ) AGNs and radio-loud (RL) AGNs~\citep{Antonucci1993, Urry1995}. The former constitutes the predominant population of AGNs, with their spectral energy distributions (SED) primarily dominated by thermal emission from accretion disks observed across optical, ultraviolet, and X-ray wavelengths. In contrast, the latter class represents only 10--15\% of AGNs~\citep{Kellermann1989}, but is characterized by powerful jets consisting of relativistically beamed outflows composed of highly magnetized plasma. The relativistic jets make the appearance of AGNs more spectacular. When viewed from large orientation angles, radio jets extend from kiloparsec (kpc) to Megaparsec (Mpc) scales, displaying a two-sided morphology centered on the nucleus. Such misaligned RLAGNs are referred to as radio galaxies. Conversely, when the jets are viewed from small angles, the apparent structure of the source becomes compact but shows intense nonthermal radiation and variability from radio to $\gamma$-rays due to a strong Doppler-boosting effect. This type of jetted AGNs is known as blazars. Both radio galaxies and blazars are further classified into subclasses based on the observed jet powers, radio morphology and the characteristics of the observed SED and emission lines~\citep{Urry1995}.

Despite the wealth of observational features seen across the entire electromagnetic spectrum, 
our understanding on the physical processes governing the accretion and ejection associated with SMBHs is still incomplete: How exactly is matter transferred from galactic scales to event horizon scales? What control the power of relativistic jets? What causes the intrinsic diversity of AGNs? One obvious challenge is that AGN and its environment are highly complex and a variety of complicated physical processes are at work.  Besides, the relevant physical scales associated with the BH accretion/ejection are tiny, typically within parsec (pc) scales or much less, making them difficult to spatially resolve with most current astronomical facilities. 
Hence, the key to advancing our understanding lies in identifying a nearby target that enable detailed spatial resolution and investigation of the SMBH vicinity, coupled with the availability of high-resolution observational facilities.

\subsection{M\,87} 
\label{ssec:m87}

In this review we specifically focus on \M87 (NGC\,4486), a nearby giant elliptical galaxy in the center of the Virgo cluster located at a distance of $D=16.8$\,Mpc~\citep{blakeslee2009, bird2010, cantiello2018}. This object was originally discovered and cataloged by Charles Messier in 1781. In 1918, Heber Curtis at Lick Observatory detected \textit{`a curious straight ray \dots apparently connected with the nucleus by a thin line of matter'} in their optical image of this galaxy~\citep{curtis1918}. Although the exact physical origin of this feature was not known at that time, this was later recognized as the first discovery of an astrophysical jet. About 40 years later, Walter Baade discovered that the light from the `jet' was strongly polarized, first indicative of nonthermal synchrotron origin of the jet emission~\citep{baade1956}. 
M\,87 has also been known as one of the brightest radio sources on the sky (as Virgo\,A or 3C\,274) since the early days of radio astronomy~\cite[e.g.,][]{bolton1949}. The large-scale radio structure associated with the galaxy extends as large as 80\,kpc~\citep{mills1952, baade1954, owen2000}, and Fanaroff and Riley classified this source as `Class I' based on its radio morphology and relatively low radio luminosity~\citep{fr1974}\footnote{Based on more than 50 samples of radio galaxies, \citet{fr1974} classified radio galaxies into Class I and Class II, which are currently known as FR-I and FR-II, respectively. FR-I radio galaxies often have symmetric radio jets whose intensity falls away from the nucleus, while FR-II radio galaxies typically exhibit more highly collimated jets leading to bright hot spots far from the nucleus. \citet{fr1974} also found that FR-I/FR-II sources have lower/higher luminosities, with a division around $\sim 2\times 10^{25}\,{\rm W Hz^{-1}str^{-1}}$ at 178\,MHz. These facts have led to the suggestion that FR-I sources host relatively weakly accreting AGNs, while FR-II sources contain more powerful AGNs.}. After 1980s the advent of the Very Large Array (VLA) revealed the kpc-to-subkpc-scale radio structures in great detail (along with a well-collimated one-sided jet) thanks to its high-dynamic-range imaging capability at subarcsecond-scale angular resolution~\cite[e.g.,][]{biretta1983, owen1989}. In 1990s, high-resolution optical spectroscopic observations with the Hubble Space Telescope (HST) revealed a rotating ionized gas disk in the central pc-scale nuclear regions of the galaxy~\citep{harms1994,ford1994,macchetto1997}, which provided the first strong hint for the presence of a central black hole with exceeding a billion solar masses in the center of this galaxy (see Sect.~\ref{sec:mass} for details).

\begin{figure}[ttt]
    \centering
    \includegraphics[width=\columnwidth]{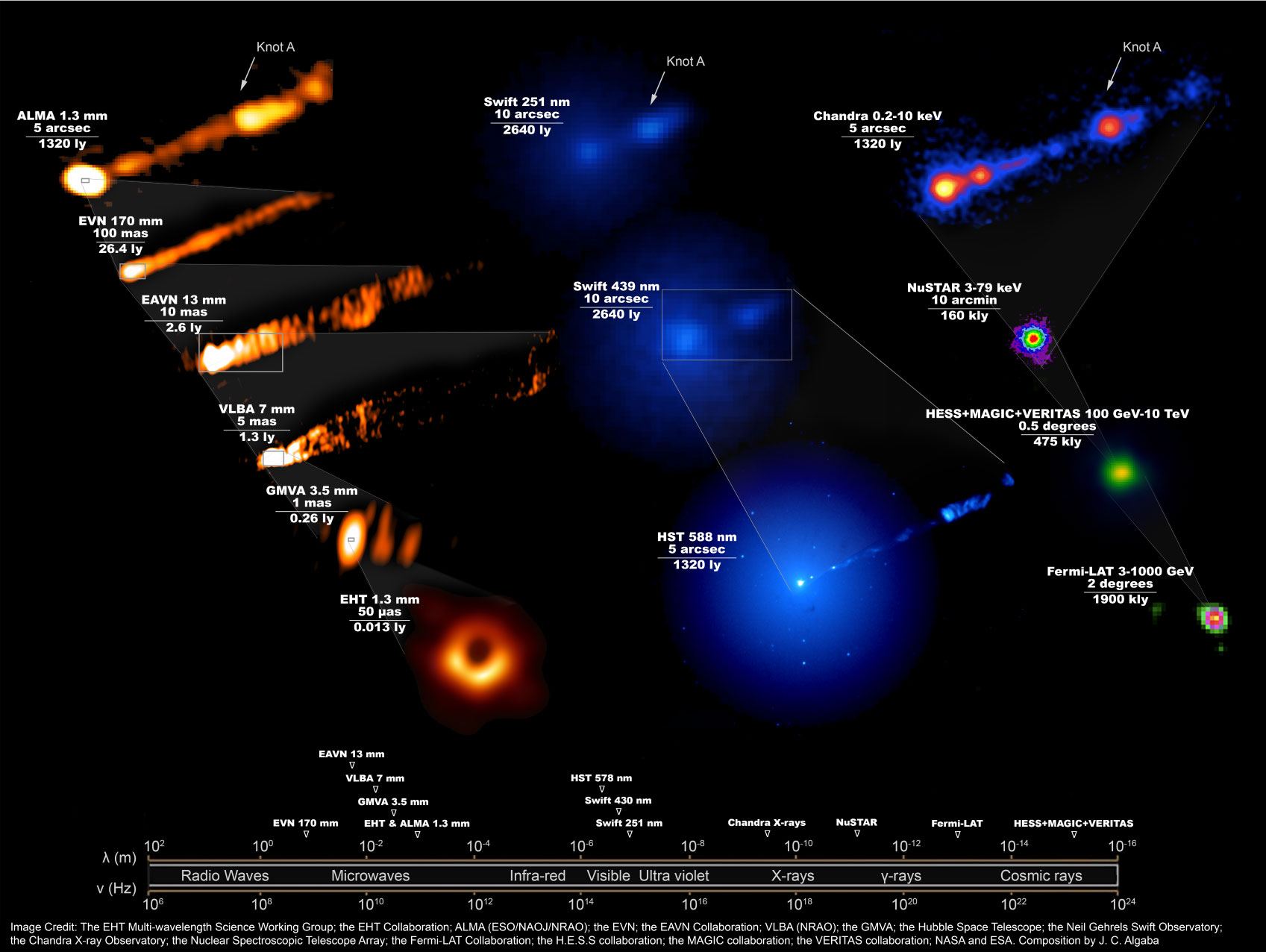}
    \caption{Broadband multi-wavelength images of M\,87~\citep{EHTMWL2021}, highlighting the energetic nature of this object across 16 orders of magnitude in the electromagnetic spectrum and over 8 orders of magnitude in spatial scales. Figure produced by courtesy of Juan Carlos Algaba} (a full image credit is described at the bottom of the figure).
    \label{fig:m87mwl}
\end{figure}

Since its discovery, \M87 has been a privileged target for studying the physics of accretion and jet formation associated with an SMBH. Compared to distant quasars, \M87 is much closer to us. Combined with a large mass of the central SMBH, this makes the apparent diameter of the SMBH second largest on the sky after the Galactic Center \SgrA, which allows us to resolve the source structure into smaller gravitational scales at a given angular resolution. On the other hand, unlike \SgrA, \M87 exhibits a powerful relativistic jet, which is the most outstanding feature characterizing RLAGNs. Moreover, \M87 is sufficiently bright across the entire electromagnetic spectrum not only in radio/optical but also in X-rays and $\gamma$-rays, making this source an excellent target to study broadband multi-wavelength (MWL) properties associated with AGN activities~(Fig.~\ref{fig:m87mwl}). 

In particular, rapid and remarkable advances have recently been made by high-resolution very-long-baseline-interferometry (VLBI) observations at radio wavelengths. The most groundbreaking example is the detection of the black hole shadow with the Event Horizon Telescope~\citep[EHT;][]{EHTC2019a,EHTC2019b,EHTC2019c,EHTC2019d,EHTC2019e,EHTC2019f}, a global VLBI network operated at 1.3 millimeters (mm). This provided the first visual evidence for the existence of SMBH as the central engine of AGN (see Sect.~\ref{sec:mass}). Global VLBI observations at mm wavelengths (including 3\,mm) are also capable of probing the inner part of accretion flows for M\,87 (see Sect.~\ref{sec:accretion}). At even longer wavelengths, VLBI observations are generally more sensitive to the larger-scale emission, enabling us to image the  collimation and acceleration regions of relativistic jets extending from near-horizon scales to galactic/kpc scales (see Sect.~\ref{sec:jet}).

\begin{figure}[ttt]
    \centering
    \includegraphics[width=\columnwidth]{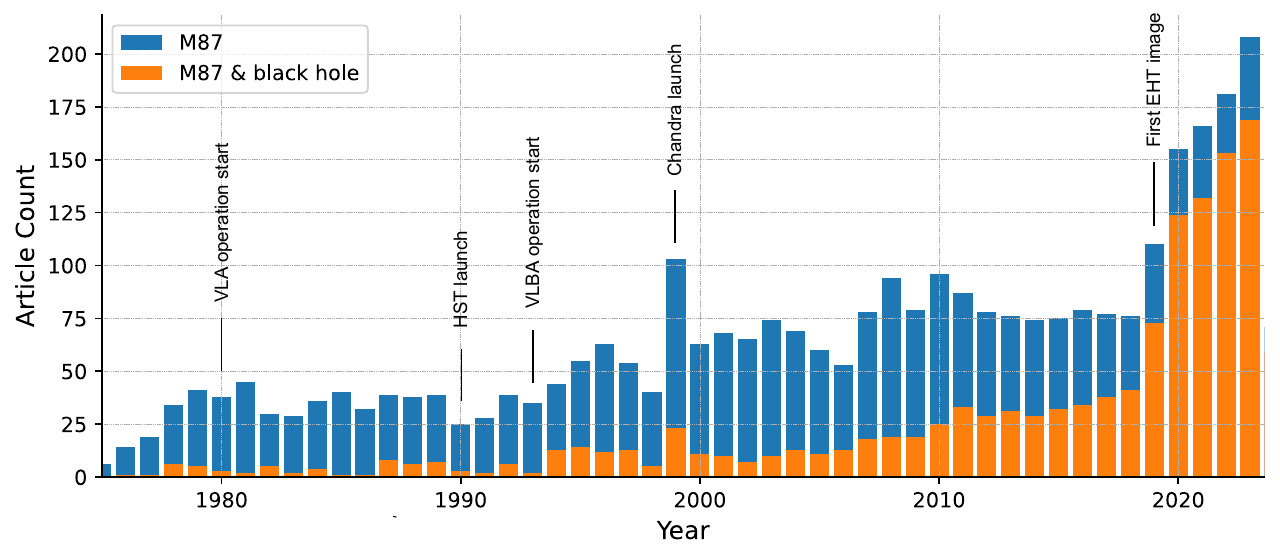}
    \caption{Annual trends in article publication: The blue-colored bar graph illustrates the total number of publications where the term `M\,87' is included somewhere in the title, abstract, or main text. The orange-colored bar graph indicates the fraction of publications where `black hole' is additionally included. The data were obtained from NASA's Astrophysics Data System.}
    \label{fig:papers}
\end{figure}

In summary, \M87 provides a rare opportunity to address various key questions relevant to (RL)AGNs as well as the physical connection among the SMBH, accretion flows and relativistic jets, especially under the low accretion/Eddington regime (see Sect.~\ref{sec:accretion}), across all electromagnetic wavelengths and spatial scales. In fact, despite being a single object, \M87 has been studied for many decades in the astrophysical community, with international workshops focused on this object~\citep[e.g.,][]{blandford1999a}. Attention to this object has been growing even more rapidly in the last few years. Fig.~\ref{fig:papers} displays annual trends in article publication related to M\,87. While the number of annual publications has  steadily increased from 1970s to 2010s, one can see a dramatic rise in publications since 2019/2020, very likely motivated by the first EHT imaging results. Therefore, now it would be a good time to review the current observational understanding of this active galaxy.

\subsection{Structure of this review}\label{ssec:strcutureofthisreview}
In this review, we provide a comprehensive overview of the current observational understanding of the active galaxy \M87, covering a broad range of spatial scales from event horizon scales to kpc scales. The review is structured as follows. In Sect.~\ref{sec:key}, we first summarize the key parameters characterizing AGN/SMBH intrinsic activity. In Sect.~\ref{sec:mass}, \ref{sec:accretion} and \ref{sec:jet}, we highlight various observational progress on M\,87, dividing the topics into three key components: the central SMBH, the accretion flows onto SMBH, and the relativistic jet, respectively. 
In Sect.~\ref{sec:wheredowestand}, we summarize our current understanding of this source and highlight some future prospects. Finally, we conclude the review in Sect.~\ref{sec:summary}. 
Readers may also refer to \citet{Boccardi2017} for a focused review on millimeter-VLBI observations of AGNs, and to \citet{blandford19} for an extensive review covering both observational and theoretical efforts on studies of AGN jets in general.

\bigskip

\section{Key parameters for characterizing AGNs}\label{sec:key}

Galaxies are fundamental building blocks in the universe, and humans live in the Milky Way galaxy. Almost every galaxy is now believed to host an SMBH at its center. 
One of the most remarkable predictions of Einstein’s theory of general relativity (GR) is the existence of BHs. In fact, recent EHT and mm-wave VLBI images of \M87 resolved the BH shadow for the first time (Sect.~\ref{sec:mass}), and the accretion flows with the emerging jet (Sect.~\ref{sec:accretion} and \ref{sec:jet}). These three key ingredients have constituted our solid hypothesis for the AGN paradigm over the past decades. 
However, we still do not fully understand even the basic property of the galactic nucleus in the Milky Way galaxy. Looking beyond our own system, is the AGN paradigm really true for galactic centers other than \M87? If so, what are the fundamental parameters governing the strength and diversity of AGNs? Thanks to the ultra-high-resolution radio imaging down to $\sim$20\,microarcseconds ($\mu$as) scales, we are now reaching the vicinity of SMBH in both our galactic center SgrA$^{\ast}$ and M\,87, which finally allows us to directly test what we have hypothesized over the past decades.
Here we summarize some key parameters in the study of SMBHs/AGNs. Extensive studies utilizing a suite of cutting-edge observational instruments allow us to quantitatively investigate these parameters in the pursuit of understanding the accretion/ejection physics.

\subsection{Black hole mass ($M_{\rm BH}$)}\label{ssec:key-mass}

SMBHs are found in most, if not all, galactic nuclei with a mass range of $M_{\rm BH}$, from $\sim$$10^{6} \, M_{\odot}$ to $\sim$$10^{10} \, M_{\odot}$~\citep{Kormendy2013}. RLAGNs exist in a variety of host galaxies, from disk-dominated spirals to giant ellipticals, with a similar $M_{\rm BH}$ range as described above \citep{ho2002}. It seems that there is no clear difference in the BH mass range ($10^{6} - 10^{10}\, M_{\odot}$) between RLAGNs and RQAGNs \citep{woo2002}, implying that $M_{\rm BH}$ may not be a dominant factor in characterizing AGN activity at radio bands.

Blazars, the other population of RLAGNs, fall into two sub-classes i) flat-spectrum radio quasars (FSRQs; objects with emission-line dominated spectra) and ii) BL Lac objects (BLOs; nearly line-less objects), with generally higher SED peaks (for both synchrotron and inverse Compton radiation) in FSRQs than in BLOs. The $M_{\rm BH}$  of blazars spans from $\sim$$10^{7.5}\, M_{\odot}$ to $\sim$$10^{9.5}\, M_{\odot}$ \citep{wu2002, shaw2012, xiao2022} with a possible localization for BLOs at $10^{8}-10^{9}\, M_{\odot}$ and for FSRQs at $10^{8}-10^{9.5}\, M_{\odot}$ \citep{ghisellini2010, chen2015}.

The majority of RLAGNs (`jetted') have been believed to be hosted by elliptical galaxies with high BH masses $M_{\rm BH} \gtrsim 10^{8}\, M_{\odot}$ \citep{fabian1988}, while late-type galaxies, such as spirals with lower BH masses $M_{\rm BH} \lesssim 10^{8}\, M_{\odot}$, were traditionally considered as RQAGNs (`non-jetted') \citep{laor2000}.
This conventional hypothesis, however, has been updated during the last two decades by discovering radio-loud narrow-line Seyfert 1 galaxies (RLNLSy1s) with $M_{\rm BH} \sim 10^{6} - 10^{8}\, M_{\odot}$ \cite[e.g.,][]{komossa2006, Yuan08, Mathur12}. Furthermore, the detection of $\gamma$-ray emitting RLNLSy1 objects \citep{Abdo09} has opened a new era for understanding the jetted RLAGNs hosted in diverse types of galaxies, including spirals, with a wide range of $M_{\rm BH} \sim 10^{6} - 10^{10}\, M_{\odot}$.

There are various independent methods proposed to calculate $M_{\rm BH}$: the traditional virial BH mass, the stellar velocity dispersion or the bulge luminosity, and the variation time scale. The $M_{\rm BH}$ of \M87 has been estimated with several methods as introduced in Sect.~\ref{sec:mass}.

\begin{figure}[ttt]
    \centering
    \includegraphics[width=12cm]{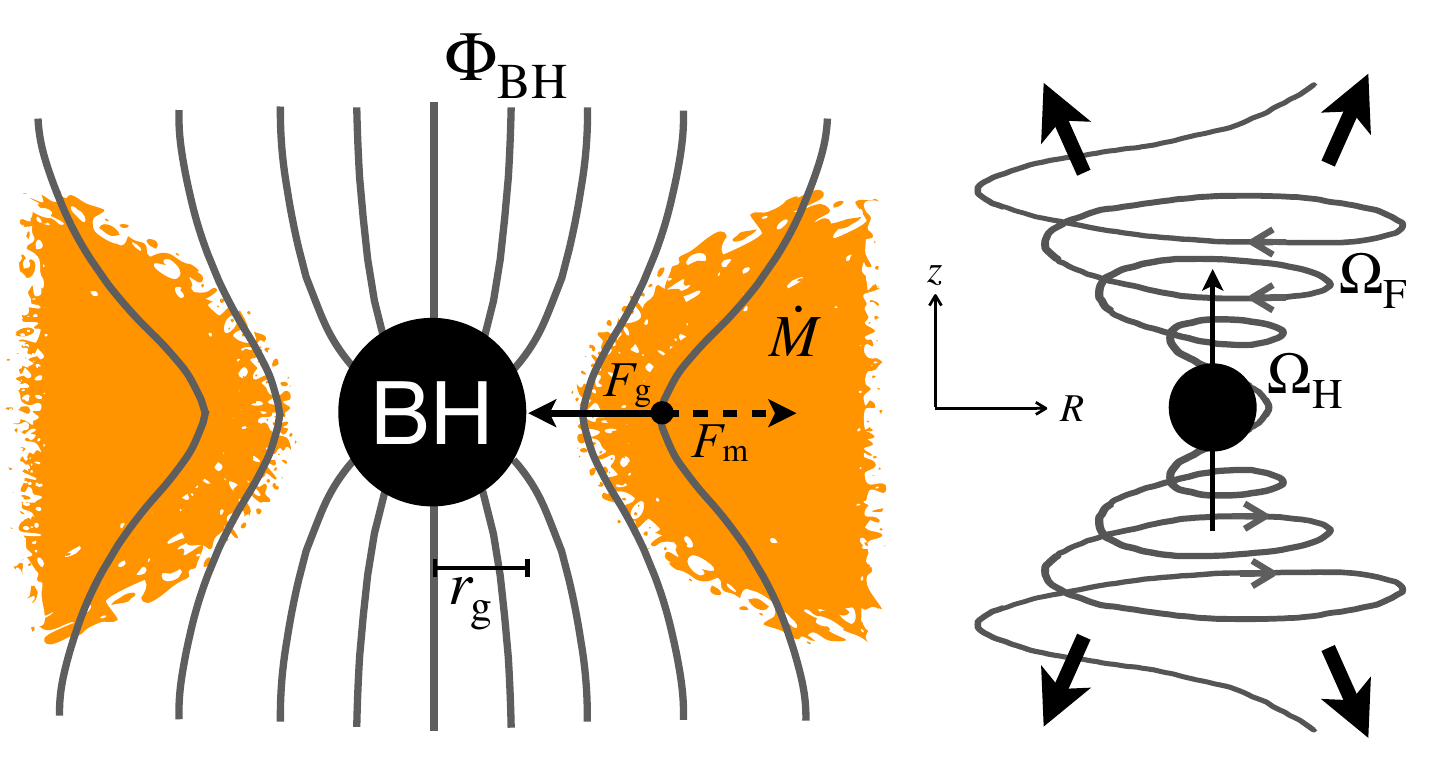}
    \caption{Schematic views of the BH accretion in radiatively-inefficient regimes with advected magnetic fluxes (Left) and the production of Poynting flux by a rotating BH (Right). Key parameters in Sect.~\ref{sec:key} are displayed. Note that both a cylindrical radius $R$ and a spherical radius $r$ are used throughout this paper.}
    \label{fig:schematic}
\end{figure}

\subsection{Black hole spin ($a$)}\label{ssec:key-spin}

BHs can rotate like other stars. Their rotation speeds could be close to the speed of light, causing a dramatic change in the metric and in the way matter moves (the mass accretion and the energy extraction). The properties of a BH are uniquely characterized by its mass, angular momentum, and electrical charge, based on the no-hair theorem of GR. In reality, a charged BH may be hard to exist due to its neutralization by discharge to its surroundings (difficult to maintain particle creation and/or accretion of matter onto the hole). Therefore, astrophysical BHs can be determined solely by the mass and angular momentum (henceforth referred to as spin).

An uncharged, rotating (spinning) BH is described by the Kerr metric \citep{kerr1963} with the BH mass $M_{\rm BH}$ and angular momentum $J$. A dimensionless angular momentum (spin) parameter ($-1 \le a \le 1$) that measures a fraction of the angular momentum $J$ relative to its maximum value is defined as follows: 
\begin{align}
a \equiv \frac{J}{(G M_{\rm BH}^{2}/c)},
\end{align}
where $G$ and $c$ are the gravitational constant and the speed of light, respectively. Here, we introduce the `gravitational radius' $r_{\rm g} \equiv G M_{\rm BH}/c^{2}$ as the standard definition. The BH can be rotating in a prograde ($a > 0$) or retrograde ($a < 0$) manner in that system.

We also do want to locate the event horizon ($r_{\rm H}$), the boundary of the BH. In the Schwarzschild metric \cite[a static metric for the spacetime around a non-rotating BH with $a=0$, where the spatial coordinates are now spherical-polar,][]{schwarzschild1916}, we have $r_{\rm H} = 2 G M_{\rm BH}/c^{2}$ 
($r_{\rm S}$: Schwarzschild radius). But for the Kerr metric, the radius of the BH event horizon is described as 
$r_{\rm H} \equiv r_{\rm g} (1 + \sqrt{1-a^{2}})$,
where $r_{\rm H} \rightarrow r_{\rm g}$ with maximally spinning cases in both prograde and retrograde fashions. Note that the horizon is still spherical even for a spinning BH. Another important radius appears in the Kerr metric,
$r_{\rm E} \equiv  r_{\rm g} (1 + \sqrt{1-a^{2} \cos^2{\theta}})$,
which is larger than $r_{\rm H}$ for all $\theta$ and $\phi$ (azimuthal and polar angles, respectively), 
except at the poles where the two are equal. 
Note that at the equator, always $r_{\rm E}=2 r_{\rm g}$, 
for any BH spin. 

The region $r_{\rm H} < r < r_{\rm E}$,
{\it called the `ergosphere'}, is peculiar:  everything must rotate in the same direction as the BH, just as everything must fall toward the BH within the horizon. The BH rotation causes the inertial frames to be dragged at an angular frequency  
$\Omega \approx \Omega_{\rm H} / (r/r_{\rm H})^{-3}$, 
where
\begin{align}\label{eq:Omega_H}
\Omega_{\rm H} =\frac{ac}{2 r_{\rm H}} = \frac{ac}{2 r_{\rm g}(1+\sqrt{1-a^{2}})}
\end{align}
is the angular frequency of the event horizon {\cite[][]{misner1973}.
This implies that the rotation rate $\Omega$ does not vanish beyond the ergosphere, but rapidly drops off outward. Thus, the {\it frame dragging} is almost concentrated near or within this region.

In the vicinity of a spinning BH, magnetic field lines advected with the accreting matter, can also rotate. When the field lines thread the event horizon, significant amounts of energy and angular momentum are extracted from the rapidly spinning BH along all field lines (see Fig.~\ref{fig:schematic}). Each field line is wound in such a way that the toroidal (azimuthal) field component is generated. An angular frequency of the field line $\Omega_{\rm F}$ will take an intermediate value between $\Omega = \Omega_{\rm H}$ and $\Omega \sim 0$ outside the ergosphere. In general, BHs are not considered to be a perfect conductor so that the field lines threading the event horizon can slip with respect to the horizon: $\Omega_{\rm F} < \Omega_{\rm H} $.
Typically, $\Omega_{\rm F} \simeq 0.5 \Omega_{\rm H}$ is considered \cite[][]{blandford77,macdonald1982}, but it depends on the geometry of the BH magnetosphere and/or the spin parameter $a$ \cite[][]{komissarov2001, penna2013, penna2015}.

\subsection{Mass accretion rate ($\dot{M}$)}\label{ssec:key-mdot}

The mass accretion rate $\dot{M} \equiv dM/dt$, i.e. the rate at which the mass of the accreting compact object increases with time, is a fundamental parameter, and we evaluate it in units of $M_{\odot}\,{\rm yr^{-1}}$ in the astrophysical context. On galactic scales, the hot X-ray emitting intergalactic medium (IGM) through quasi-spherical accretion \citep{Bondi1952} is widely used to estimate the amount of the accretion power by the SMBH. The Bondi accretion radius $r_{\rm B}$ where the gravitational potential of the SMBH dominates over the thermal energy of the IGM gas, is given by
\begin{align}
    r_{\rm B} = \frac{2G M_{\rm BH}}{c_{\rm s}^{2}} \simeq 30\, {\rm pc} \left(\frac{k_{\rm B} T}{{\rm keV}}\right)^{-1} \left(\frac{M_{\rm BH}}{10^{9}\,M_{\odot}}\right),
\end{align}
where $c_{\rm s}$ and $k_{\rm B}$ are the adiabatic sound speed of the gas at the accretion radius \citep{allen06} and the Boltzmann constant, respectively. Typical values of the IGM temperature $T$ and $M_{\rm BH}$ are adopted. Correspondingly, the Bondi accretion rate 
$\dot{M}_{\rm B}$ is given by 
\begin{align}
    \dot{M}_{\rm B} = \frac{\pi \rho G^{2} M_{\rm BH}^{2}}{c_{\rm s}^{3}}
    \simeq 0.012\, M_{\odot}\,{\rm yr^{-1}} \left(\frac{n_{\rm e}}{{\rm cm}^{-3}}\right) \left(\frac{k_{\rm B} T}{{\rm keV}}\right)^{-3/2} \left(\frac{M_{\rm BH}}{10^{9}\,M_{\odot}}\right)^{2}
\end{align}
for an adiabatic index $\gamma=5/3$ and the mass density $\rho$ with the electron number density $n_{\rm e}$ \citep{russell13}.

Suppose that the BH accretion starts around the sphere of influence ($\simeq r_{\rm B}$), where the BH's gravity becomes dominant relative to that of the host galaxy, at an accretion rate similar to the Bondi rate $\simeq \epsilon_{\rm acc} \dot{M}_{\rm B}$ (where $\epsilon_{\rm acc}\leq 1$ is the accretion efficiency). For the Bondi accretion of M\,87, readers can refer to Sect.~\ref{ssec:accretion_bondi}. 
In the radiatively inefficient regime, we assume a typical value of the viscosity parameter $\alpha \sim 0.1-0.3$ in the hot, advection-dominated accretion flows \cite[ADAF: $\rho (r) \propto r^{-3/2}$,][]{narayan1995, mahadevan1997} and we can refer to the solution of a `giant' ADAF which provides $\epsilon_{\rm acc} \gtrsim 0.5$ \citep{narayan2011}. A flatter radial mass density profile (than the ADAF solution) $\rho (r) \propto r^{-3/2+s}\,(0 < s \le 1)$ suggests a reduction in the mass accretion rate towards the BH: $\dot{M}(r) \propto r^{s}$ (while $\dot{M}(r) = const.$ in ADAF). It would be worth considering that the hot gas is convectively unstable \citep{abramowicz2002}, moderately magnetized, and/or the outflow (wind) co-exists \citep{blandford1999}, as examined by both hydro and magnetohydrodynamic (MHD) simulations \cite[e.g.,][references therein]{yuan2014}. 

An extensive numerical survey of slowly rotating magnetized BH accretion through the Bondi radius provides a power-law slope of $s \simeq 0.2 - 0.5$ \cite[][]{pang2011}. Indeed, inside the Bondi radius, a self-consistent profile of the electron number density is obtained in X-ray observations toward \M87 (see Sect.~\ref{sec:accretion}). Three-dimensional general relativistic magnetohydrodynamics (GRMHD) simulations provide $\dot{M} \propto r^{0.5}$ at $r \lesssim 100\, r_{\rm S}$ \citep{sadowski2013, yuan2015}. Considering theoretical/numerical and direct observational results, a substantial reduction of the mass accretion rate compared with the Bondi rate would be expected towards the inner scales.

In the radiatively inefficient accretion flows (RIAFs), the mass accretion rate relative to the Eddingtion rate, $\dot{m}=\dot{M}/\dot{M}_{\rm Edd}$ is well below the unity, where 
\begin{align} \label{eq:mdot_edd}
\dot{M}_{\rm Edd} \simeq 22\, M_{\odot}\, {\rm yr}^{-1} \left(\frac{\epsilon_{\rm rad}}{0.1}\right)^{-1} \left(\frac{M_{\rm BH}}{10^{9}\, M_{\odot}}\right)    
\end{align}
with the standard radiative efficiency $\epsilon_{\rm rad}$.
There is a critical value (upper limit) $\dot{m}_{\rm crit} \sim \alpha^{2} \sim (0.01-0.1)$ for existing an ADAF \citep{narayan1995}. Given that the Bondi accretion rate is well below the Eddington rate (see Sect.~\ref{ssec:accretion_bondi}), the BH accretion in \M87 is certainly in the RIAF regime, where the disk is optically thin and hot, and thus becomes geometrically thick \cite[the ion is heated up to the virial temperature $T_{\rm i} \lesssim 10^{12}\, {\rm K}$, while the electron temperature is much cooler $T_{\rm e} \sim 10^{9-11}\, {\rm K}$:][]{nakamura1997, quataert1999}. 

In the quiescent state (QS) far below the critical mass accretion rate (say $\dot{m} < 10^{-3}$), an ADAF-type accretion flow occupies all the way down to the innermost 
stable circular orbit (ISCO) around the BH. However, there is a state transition from an ADAF to a radiatively efficient (luminous), geometrically thin, standard accretion disk \citep{shakura1973, novikov1973}. A large transition radius $R_{\rm tr} \gtrsim 10^{4}\, r_{\rm S}$ is expected in the QS such as \M87 and Sgr$\,$A$^{\ast}$ \citep{narayan2008}. In the low state (LS) where $\dot{m}$ is higher than that in QS (i.e.,  $\dot{m} \sim 10^{-3}-10^{-1.5}$), the radiative efficiency becomes larger, and the transition radius gets smaller. In the intermediate state (IS), where $\dot{m} \sim (10^{-1.5} - 10^{-1}) \lesssim \dot{m}_{\rm crit}$, the radiative efficiency is as high as $\sim 0.1$ and the transition radius approaches the ISCO. Finally, the state transition is completed in the high state (HS), where $\dot{m} \gtrsim \dot{m}_{\rm crit}$, and the accretion flow changes completely to a thin disk down to the ISCO. This system is in the thermal state so that its spectrum is well-described by a multi-color disk model
\citep[see also][for more details]{esin1997, narayan2008, ho2009}.

Among the jetted RLAGNs, low power radio galaxies such as FR-I sources (e.g., M\,87) can fall into the QS. High power radio galaxies such as FR-II sources (e.g., Cygnus A) can be in the LS. \citet[][]{ghisellini2010} studied all blazars of known redshift detected by the {\it Fermi} satellite and provided a possible `split' between BLOs and FSRQs in the accretion mode, which occurs at $\dot{m} \sim \dot{m}_{\rm crit}$ \citep[see also][]{xu2009, chen2015}. Based on these observational facts, we can categorize BLOs in the LS and the IS, while FSRQs in the HS. Furthermore, some {\it Fermi} FSRQs may be in the very high state (VHS) where $\dot{m} \gtrsim 1$. This is the so-called the `super-Eddington' accretion regime in the optically thick RIAF or slim disk \citep{begelman1982, abramowicz1988}. Also, NLSy1s are considered to be in the VHS \citep{colin2004}; transient and highly variable NLSy1s may indicate that such high-amplitude variability is linked with a transition between a standard disk and a slim disk as a result of the thermal instability, with $3 \lesssim \dot{m} \lesssim 300$ \citep{Kawaguchi2003}.

\subsection{Magnetic flux threading a black hole 
($\Phi_{\rm BH}$)}\label{ssec:key-mag-flux}

The magnetic field that an isolated BH can possess is weak: if there is a strong magnetic field, the mass accretion is prevented by its magnetic pressure and torque. 
\citet{blandford77} estimated 
the critical field strength for a maximally spinning BH as $\approx 2.5 \times 10^{5}\ {\rm G} \ (M_{\rm BH}/M_{\odot})^{-1/2}$. 
On the other hand, the field strength brought in by accreting matter can be much stronger. Thus, an intrinsic magnetic field of the BH can be considered as astrophysically unimportant. 

GRMHD simulations reveal that the poloidal magnetic flux in the accretion flow is advected inward 
until the ram pressure of accreting material ($\lesssim \rho c^{2}=G M_{\rm BH} \rho / r_{\rm g}$, shown as $F_{\rm g}$ in Fig.~\ref{fig:schematic}) is balanced with the magnetic pressure ($B^{2}/8\pi$, where $B$ is the strength of magnetic field lines, shown as $F_{\rm m}$ in Fig.~\ref{fig:schematic}).
The accumulated poloidal magnetic flux onto the event horizon
threading (one hemisphere of) 
the BH ($\Phi_{\rm BH}$)
is a key parameter that essentially characterizes the accretion flow properties. 
This is given in the following form 
\citep{Tchekhovskoy11, yuan2014}:
\begin{align}\label{eq:PhiBH}
    \Phi_{\rm BH} \equiv \phi_{\rm BH}
    (\dot{M}r_{\rm g}^{2}c)^{1/2}.
\end{align}
When
the dimensionless parameter
$\phi_{\rm BH} \gtrsim 50$
(in Gaussian units), 
a strong magnetic field pushes aside the accretion flow and tends to produce outflows. This is known as the Magnetically Arrested Disk (MAD) state
\citep[]{Igumenshchev03,Narayan03,Tchekhovskoy11}.
On the other hand,
the case $\phi_{\rm BH} \sim a~few$
corresponds to the Standard Accretion and Normal Evolution (SANE) state, where weak and turbulent magnetic fields dominate the accretion flow \citep[]{Narayan12}. \\

\subsection{Jet power ($L_{\rm j}$)}\label{ssec:key-lj}

The jet launching mechanism in AGNs is a longstanding challenge in astrophysics. 
Electromagnetic radiation emitted from AGN jets
spanning from radio to very-high-energy (VHE) $\gamma$ rays
have improved our understanding of AGN jets.
The total power of the jet ($L_{\rm j}$) is the sum of the 
several components as described by
\begin{align}
    L_{\rm j}=L_{\rm rad}+L_{\rm p}+L_{\rm e^{\pm}}+L_{\rm Poy}
\end{align}
where
$L_{\rm rad}$,
$L_{\rm p}$,
$L_{\rm e^{\pm}}$, and
$L_{\rm Poy}$,
are 
the radiation power,
the proton power,
the electron-positron pair power, and
the Poynting power of the jet,
respectively.
\citep[e.g.,][]{sikora20}.

If the Blandford and Znajek (BZ) process \citep{blandford77}, where the electromagnetic energy extraction from a rotating BH is operated as a jet launching mechanism,
$L_{\rm Poy}=L_{\rm BZ}$ holds at the jet base and 
BZ power ($L_{\rm BZ}$) can be given by
\begin{align} \label{eq:L_BZ}
    L_{\rm BZ} = 
    \frac{\kappa}{4\pi c} 
    \Omega_{\rm H}^{2}\Phi_{\rm BH}^{2} f(\Omega_{\rm H}),
\end{align}
where $ f(\Omega_{\rm H})=1
+ 1.38(\Omega_{\rm H}r_{\rm g}/c)^{2}
- 9.2(\Omega_{\rm H}r_{\rm g}/c)^{4}$, 
$\kappa$ is the geometrical factor, and $\Omega_{F}=\Omega_{H}/2$ is assumed
\citep[see][for details]{Tchekhovskoy10,Tchekhovskoy11}.

Observed AGN luminosities form two separated sequences on the radio-loudness–Eddington-ratio plane,
suggesting larger BH spins for radio-loud sources \citep[]{sikora07}.
\citet{Zamaninasab14} suggest dynamically important magnetic fields near accreting BH based on the estimation of $\Phi_{\rm BH}$.
They seem to support the idea of the BZ process at work 
and thus, most of theoretical studies assume that magnetic fields play a key role in jet formation.
However, it is not obvious whether observed electromagnetic signals indeed suggest that the magnetic field plays a major role in the dynamics or not. 
This assertion should be verified through comprehensive observations, and M\,87 can provide us an excellent platform to explore this (see Sect.~\ref{ssec:B}).

The $L_{\rm rad}$ is directly obtained 
from observational data and
they are mostly dominated 
by the power of synchrotron emission ($L_{\rm syn}$) and
inverse-Compton emission ($L_{\rm IC}$)
\citep[]{kataoka05}.
Based on the works of 
\citep[]{sikora07,ghisellini14}
the observed bolometric luminosity 
is known to be in a wide range of
$10^{40}~{\rm erg~s^{-1}}
\lesssim L_{\rm rad} \lesssim 
10^{48}~{\rm erg~s^{-1}}$.
On the other hand, constraining the plasma composition 
and estimating $L_{\rm p}$, and $L_{\rm e^{\pm}}$ 
in AGN jets remains challenging and is still a subject of debate since the observed emissions predominantly originated from only nonthermal $e^{\pm}$ pair
\citep[]{wardle98, sikora00, kataoka08, kino12,kawakatu16, sikora20}.

Although the estimation of $L_{\rm j}$ contains a certain degree of ambiguity due to the reason mentioned above,
the jet power of FSRQs can maximally reach up to
$L_{\rm j}\sim 10^{49}~{\rm erg~s^{-1}}$,
while that of radio galaxies
can reach as low as 
$L_{\rm j}\sim 10^{42}~{\rm erg~s^{-1}}$
\citep[see for example][and references therein]{rawlings91, daly19, chen23}.


\bigskip

\section{Central supermassive black hole}\label{sec:mass}

\subsection{BH mass}\label{ssec:mass}

The mass of the central SMBH ($M_{\rm BH}$) of \M87 has been estimated using three different approaches: stellar dynamics modeling \citep[e.g., ][]{gebhardt2011}, gas dynamical modeling \citep[e.g., ][]{harms1994, macchetto1997, Walsh13}, and BH shadow measurements \citep{EHTC2019a, EHTC2019f, EHTC2024}.

Stellar dynamical measurements involve assessing the central rise in stellar velocity dispersion towards the center of the galaxy. 
This method began with the assumption of a simple Gaussian distribution of stellar dynamics, then it has evolved to estimate the mass-to-distance ratio, $M/D$, with high accuracy through the development of sophisticated modeling techniques utilizing the full line-of-sight stellar velocity distributions achieved with high angular resolution spectroscopic observations. 
As a byproduct, it provides the mass-to-light ratio, $M/L$, of the stellar populations at its center, which can be used to validate measurements by comparing them with independent estimates derived from population synthesis. 
Utilizing these approaches, $M_{\rm BH}$ of \M87 was estimated to be $M_{\rm BH} = (6.6\pm 0.4)\times 10^{9}\,M_{\odot}$ for an assumed distance of 17.9\,Mpc (corresponding to $6.2\times 10^{9}\,M_{\odot}$ for 16.8\,Mpc) \citep{gebhardt2011}.  
Recently, new estimates on $M_{\rm BH}$ have been proposed by adopting new triaxial orbit models, resulting in a relatively smaller mass of 
($5.3_{-0.22}^{+0.37})\times10^{9}\,M_{\odot}$ \citep{Liepold23}.  
On the other hand, a larger mass of $8.7\times10^{9}\,M_{\odot}$ has been suggested by new measurements using the Canada--France--Hawaii Telescope (CFHT) and the Very Large Telescope (VLT) adaptive optics observations \citep{Simon24}.
However they also estimated a smaller mass of $5.5\times10^{9}\,M_{\odot}$ depending on the adopted stellar density profile \citep{Simon24}.
 
Gas dynamical estimates rely on modeling the dynamics of the H$\alpha$ and [N\,II] emission lines from the central gas disk. 
In early 1990s, \citet{harms1994} detected a spatially resolved ionized gas disk with the HST. 
They identified an asymmetry in the velocity field and estimated $M_{\rm BH} = (2.6\pm0.8)\times10^{9}\,M_{\odot}$. 
Following this, \citet{macchetto1997} revealed the rotation curve of the ionized gas disk surrounding the nucleus. 
The velocities measured from the emission lines were consistent with the hypothesis that the ionized gas disk undergoes Keplerian rotation, leading to  
$M_{\rm BH} = (3.6 \pm 1.0) \times 10^{9}\,M_{\odot}$. 
Later, \citet{Walsh13} further refined the analysis and derived $M_{\rm BH} = (3.3 \pm 0.8) \times 10^{9}\,M_{\odot}$ 
based on comprehensive gas-dynamical modeling with new HST data acquired with the Space Telescope Imaging Spectrograph (STIS).
Historically, these $M_{\rm BH}$ estimates using gas dynamics have shown excellent agreement with each other, whereas there has been a factor of $\sim$2 discrepancy with the $M_{\rm BH}$ derived from the stellar dynamics.

While the above two methods are based on the orbital motion of gas or stars relatively far from the central BH, it is also possible to estimate $M_{\rm BH}$ by directly measuring the size of the BH shadow using ultra-high-angular-resolution VLBI observations at millimeter wavelengths. According to GR, the apparent size of the BH shadow is expected to be approximately 5 times the Schwarzschild radius \citep[e.g., ][]{EHTC2019f}. The first observational hint for such a gravitationally-lensed shadow feature was obtained for Sgr\,A* and M\,87, with proto-EHT experiments at 1.3\,mm using three stations in the United States \citep{Doeleman08, Doeleman12}. The first-ever BH shadow image of \M87 obtained with EHT discovered an asymmetric ring-like structure with a diameter of 42 $\pm$ 3 $\mu$as \citep[][Fig.~\ref{fig:black hole shadow} left]{EHTC2019a, EHTC2019d}.

Comparing the observed M\,87 EHT images with a large GRMHD simulation library, they concluded that the observed ring-like structure is in good agreement with the shadow of a spinning Kerr BH \citep{EHTC2019e}. By calibrating the observed ring diameter with the predicted ring diameter based on the simulations, they derived the gravitational radius and estimated $M_{\rm BH}$ to be $M_{\rm BH} = (6.5 \pm 0.7) \times 10^{9}\,M_{\odot}$ assuming a distance of 16.8 Mpc \citep{EHTC2019f}. 
The first EHT results based on the 2017 observations have been confirmed with another independent EHT dataset taken in 2018 \citep[][Fig.~\ref{fig:black hole shadow} right]{EHTC2024}, providing strong supporting evidence that the ring seen in the EHT images represents the BH shadow predicted by GR.

The $M_{\rm BH}$ estimated from the EHT data is in good agreement with that derived from the stellar dynamics, while it deviates from the ones estimated from the gas dynamics with a confidence level of 99 $\%$ \citep{EHTC2019f}. This discrepancy may be attributed to the inclusion of internal gas motions within the disk, along with adopting slight changes in the inclination angle of the disk \citep{Jeter19, Jeter21}. Therefore a further study is warranted, particularly using future high-resolution optical instruments.

\begin{figure}[ttt]
    \centering
    \includegraphics[width=\columnwidth]{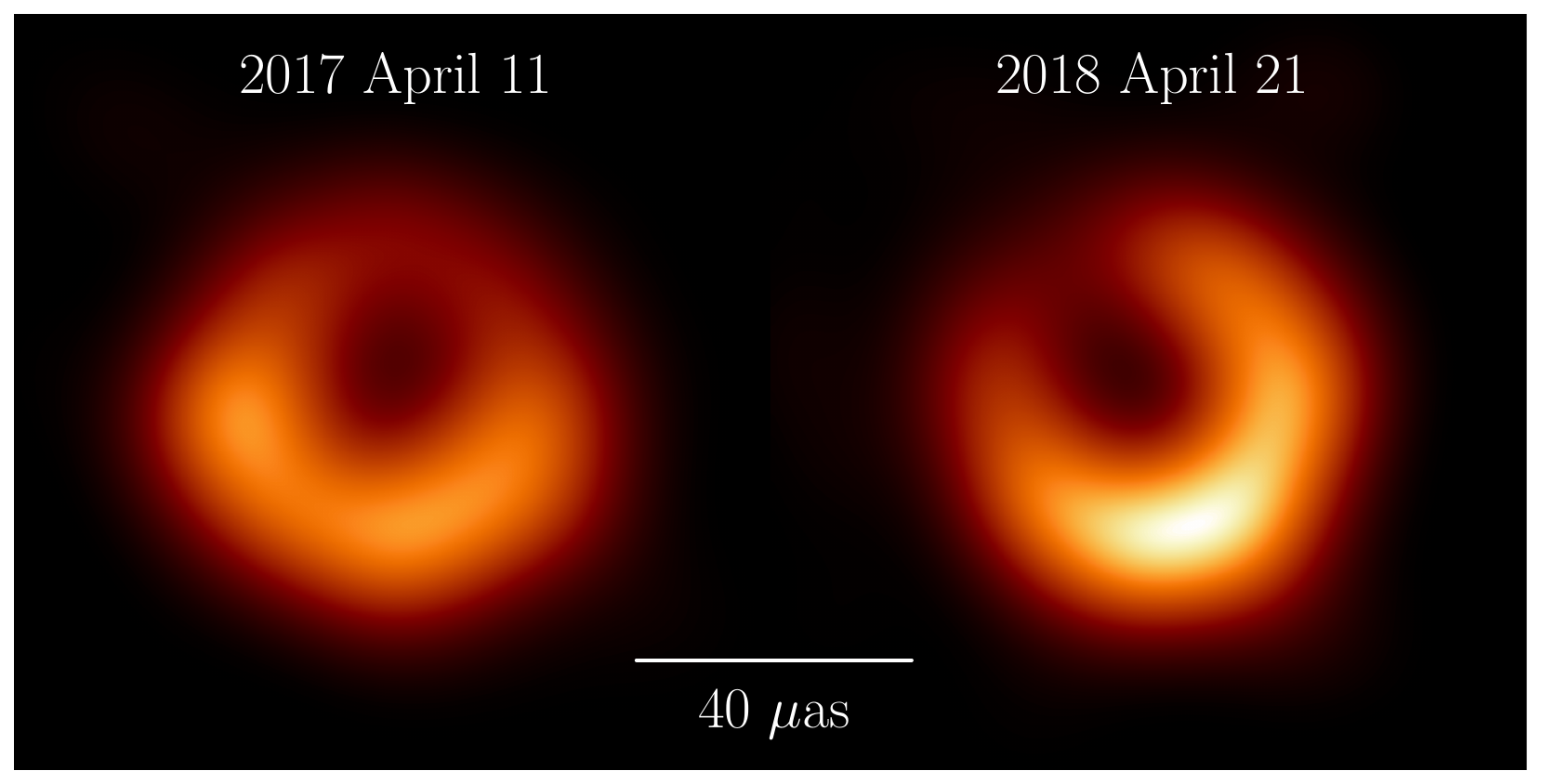}
    \caption{The first-ever image of a black hole shadow, captured by EHT in 2017 (left), and its appearance one year later in 2018 (right). The comparison of the two images shows that the overall characteristic features, such as the ring size and the brighter southern emission,  remain consistent over a year.
    \citep{EHTC2019a, EHTC2024}. Figure credit: EHT Collaboration}
    \label{fig:black hole shadow}
\end{figure}

\subsection{BH spin}\label{ssec:spin}

The spin of a BH is a fundamental parameter and crucial for describing its nature. In astrophysical contexts, the BH spin emerges as a compelling explanation for the extraordinary power exhibited by jets. Take, for example, the case of M\,87, where the estimated mass accretion rate onto the SMBH ranges from 10$^{-3}$ to 10$^{-4}$\,$M_{\odot}\,{\rm yr}^{-1}$. This translates to a mass accretion power $\dot{M} c^{2}$ of $\sim$10$^{43-44}$\,erg\,s$^{-1}$ (as discussed in Sect.~\ref{ssec:accretion_horizon}). However, the expected kinematic power of the jets falls in the range of $\sim$10$^{43-45}$\,erg\,s$^{-1}$ (as outlined in Sect.~\ref{ssec:Lj}).  Hence, it requires an energy conversion efficiency of $\gtrsim 100 \%$, implying that the simple mass accretion power alone may be insufficient to explain the jet power. It is postulated that the subtraction of the rotating energy of the BH can compensate for this difference. 

The direct imaging of SMBH shadow provides another important clue regarding the nature of BH beyond its mass. 
The observed brightness asymmetry of the 1.3\,mm ring is likely caused by the effect of Doppler beaming associated with the frame dragging effect of a spinning Kerr BH. Since the observed ring appears brighter in the southern part of the emission, it suggests that the BH spin vector is oriented away from Earth if the BH spin axis aligns with the large-scale jet \citep{EHTC2019f}.  The bright spot observed in 2018 shifted by 30 degrees relative to that observed in 2017, and the asymmetry observed in 2018 is more consistent with the orientation of the large-scale jet \citep{EHTC2024}.  These observed features indicate the presence of a spinning BH at the center of M\,87, although the value of the spin has not yet been tightly constrained in these observations. Further strong evidence of the spinning black hole will be provided by precise measurement of the photon ring with next generation VLBI observations (see Sect.~\ref{ssec:nextstep}), 
and the next step is to determine the direction of the spin relative to the rotation of the accreting matter (prograde/retrograde).

\bigskip

\section{Accretion flows}\label{sec:accretion}
\subsection{Onset of BH accretion: the Bondi scales}\label{ssec:accretion_bondi}

As described in Sect.~\ref{ssec:key-mdot}, the nature of accretion flows onto SMBH can be probed by observing the X-ray hot gas atmosphere.  
The density and temperature profiles of the X-ray hot gas in \M87 on kpc scales were extensively investigated using Chandra X-ray data \citep{DiMatteo03, russell18}. \citet{DiMatteo03} inferred the decrease in the gas temperature from 10\,kpc down to the central 1\,kpc. Within this distance range, they observed the density profile to flatten inside the central 2\,kpc, while it follows $r^{-1}$ beyond 2\,kpc up to 10\,kpc (see Fig.~\ref{fig:DiMatteo1}). 
With these observed profiles, they calculated the Bondi radius to be $\sim$150\,pc for a BH mass of $3\times 10^{9}$\,$M_{\odot}$ and estimated the Bondi accretion rate to be $\dot{M}_{\rm B} \sim 0.1\,M_{\odot}\,\rm{yr}^{-1}$.
The temperature and density profiles were further refined by \citet{russell18} through a stacking analysis of twelve 5-ks Chandra data. 
By adopting the refined temperature and density profile together with a newly estimated $M_{\rm BH}$ of 6.5 $\times$ 10$^{9}$ \msun{} \citep{EHTC2019a, EHTC2019f}, we estimate the Bondi radius and Bondi accretion rate to be 

\begin{align}
r_{\rm B} \simeq 220\, {\rm pc} \left(\frac{k T_{\rm B}}{0.9\, {\rm keV}}\right)^{-1} \left(\frac{M_{\rm BH}}{6.5 \times 10^{9}\, M_{\odot}}\right),    
\end{align}
and
\begin{align}
\dot{M}_{\rm B} \simeq 0.22\, M_{\odot}\,{\rm yr}^{-1} \left(\frac{n_{\rm e}}{0.3\, {\rm cm}^{-3}}\right)   \left(\frac{k T_{\rm B}}{0.9\, {\rm keV}}\right)^{-3/2} \left(\frac{M_{\rm BH}}{6.5 \times 10^{9}\, M_{\odot}}\right)^{2}.
\end{align}

\begin{figure}[ttt]
    \centering
    \includegraphics[width=0.49\columnwidth]{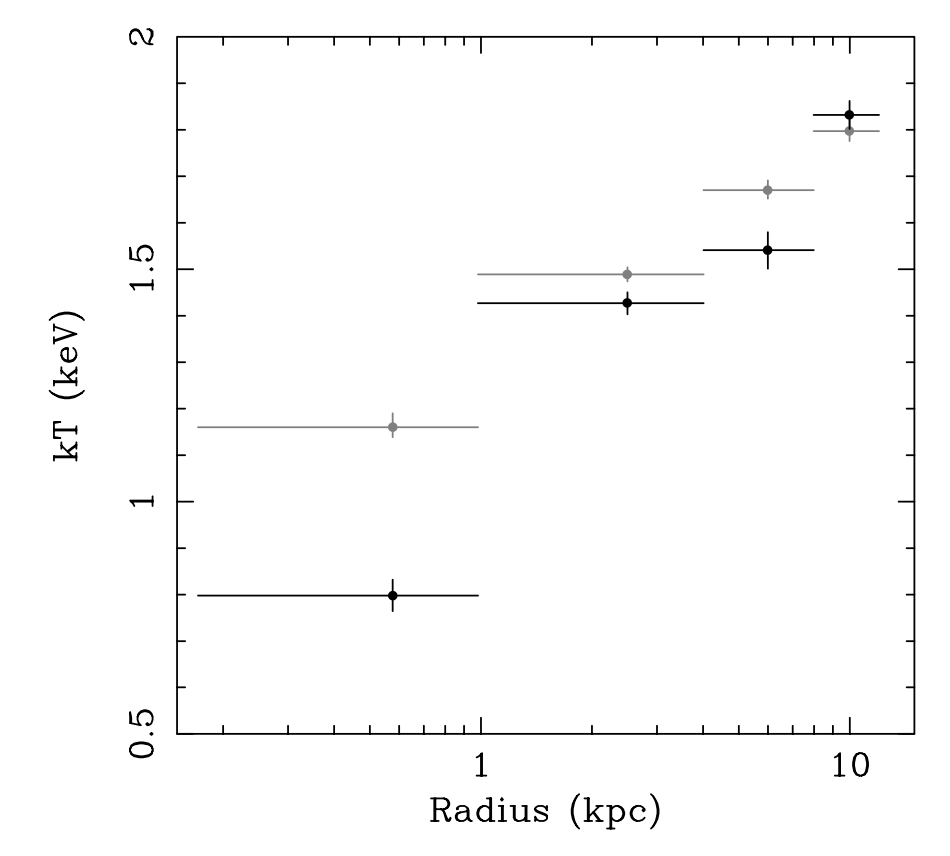}
    \includegraphics[width=0.49\columnwidth]{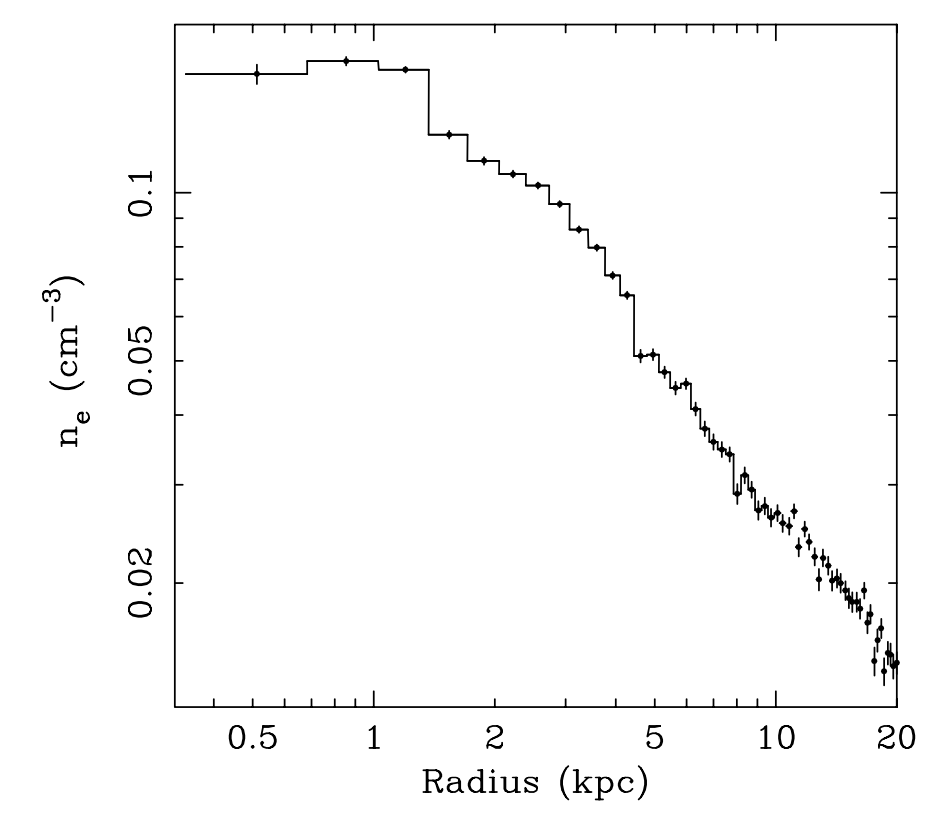}
    \caption{Temperature and density profiles of the X-ray hot gas derived by Chandra observations. A clear plateau is seen in the density profile. Images reproduced with permission  from \citet{DiMatteo03}, copyright by the AAS.}
    \label{fig:DiMatteo1}
\end{figure}

Hence, the comparison with $\dot{M}_{\rm Edd}$ introduced in Eq.(\ref{eq:mdot_edd}) ($\simeq140\,M_{\odot}\,{\rm yr^{-1}}$ for $\epsilon_{\rm rad}=0.1$ and $M_{\rm BH}=6.5\times 10^9\,M_{\odot}$) indicates that the accretion onto the M\,87 nucleus is significantly sub-Eddington, as also inferred from the underluminous core X-ray luminosity \citep[$\sim 7 \times 10^{40}$\,erg\,s$^{-1}$;][]{DiMatteo03}. 
It is notable that \citet{russell18} found no evidence of gas temperature increase within a radius of 0.25\,kpc, as expected in the classical Bondi-type accretion flows, while a further increase in density profile was observed within 0.3\,kpc. This led them to suggest that the classical Bondi accretion flow may not be accurately achieved, and that a substantial decrease in the mass accretion rate onto the central SMBH within $r_{\rm B}$.

In addition to probing the hot plasma through X-ray observations, efforts have been made to unveil the cold accretion flow at similar or slightly smaller scales using radio interferometry data. \citet{Tan08} utilized the Submillimeter Array (SMA) at 230\,GHz to detect CO (J = 2–1) line emission, aiming to identify the presence of cold molecular gas in a thin disk around the M\,87 BH. Although they detected 230\,GHz continuum emission from the nucleus and several knots at arcsecond scals, they estimated a conservative upper limit on the mass of molecular gas to be $\sim$ 8 $\times$ 10$^{6}$\,$M_{\odot}$ within 100\,pc of the central BH.  Subsequently, \citet{Simionescu18} detected extended CO (2–1) line emission approximately 40 arcseconds away, corresponding to 3.4 kpc in the southeast direction from the nucleus, using the Atacama Large Millimeter/submillimeter Array (ALMA). They derived the corresponding molecular gas mass to be (4.7 $\pm$ 0.4) $\times$ 10$^{5}$ \msun{}. Following this, \citet{Li22} conducted a systematic survey utilizing ALMA archival data. However, they did not find conclusive evidence of CO line emission in the vicinity of the M\,87 nucleus.

\subsection{Inside the Bondi radius}\label{ssec:accretion_withinbondi}

It is difficult for the current X-ray instruments to spatially resolve the scales well below $r_{\rm B}$. To explore the nature of accretion flows within $r_{\rm B}$ scales, a powerful approach would be the measurement of polarization and Faraday rotation measure (RM) at mm/submm wavelengths since the bulk of mm/submm synchrotron emission from ADAF/RIAF-type accretion flows is likely originated in the close vicinity of the central SMBH~\citep[e.g.,][]{yuan2014}. This approach was originally developed for the nearest SMBH \SgrA, which exhibits highly linear polarization with significant RM \citep[e.g., ][]{Aitken00, Bower03, Marrone06, Marrone07}. 
By attributing the observed RM to the magnetized plasma associated with the accretion flow, the accretion rate of Sgr\,A$^*$ at small radii was constrained to be $\dot{M} \sim10^{-8}\,M_{\odot}\,\rm{yr}^{-1}$ depending on the adopted accretion models \citep{Marrone06, Marrone07}.

Following the success in constraining the $\dot{M}$ of \SgrA, \citet{Kuo14} applied a similar method to \M87 using 1.3\,mm polarimetric data taken by the SMA. 
They obtained an upper limit of $\lvert{\rm RM}\rvert$ (the magnitude of RM) to be $7.5 \times 10^{5}\, {\rm rad\, m}^{-2}$. Adopting a simple analytic model of the accretion flow similar to the case of \SgrA, the $\dot{M}$ onto the \M87 SMBH was constrained to be $< 9 \times 10^{-4}\,M_{\odot}\,\rm{yr}^{-1}$. This indicates a substantial decrease in $\dot{M}$ compared to the Bondi accretion rate. This is in good agreement with expectations from the RIAF models, ruling out the classcial ADAF.  

Large RM values towards the \M87 nucleus were later confirmed by \citet{Goddi21} using multi-epoch ALMA data at 3 and 1.3\,mm. Thanks to the improved sensitivity of ALMA, they further detected significant time variation of RM both in magnitude and sign. This suggests a more complicated origin of the Faraday rotation rather than a simple accretion flow, making a more accurate estimate of $\dot{M}$ rather challenging. 

To look into the structure inside $r_{\rm B}$ more directly, high-resolution VLBI observations are required. Using the Very Long Baseline Array (VLBA) at cm wavelengths, there were several attempts to explore the pc-scale polarimetric and RM properties of \M87~\citep[e.g.,][see also Sect.~\ref{ssec:polarization}]{Junor2001, Zavala2002, Park2019b}. Most notably, \citet{Park2019b} revealed a spatially-resolved RM profile inside $r_{\rm B}$ and found that $\lvert {\rm RM} \rvert$ decreases with distance from 10\,mas to 350\,mas (corresponding to deprojected distances 5000--200000\,$r_{\rm S}$) away from the nucleus. They indicated that the observed radial slope of RM at these scales is reproduced by a gas density profile of $\rho(r) \propto r^{-1}$ for the Faraday screen. Such a density profile is expected for RIAF-type hot accretion flows with substantial winds or mass loss~\citep[ADiabatic Inflow-Outflow Solutions (ADIOS);][]{blandford1999}, which is consistent with the significant reduction of $\dot{M}$ near SMBH inferred from the mm/submm interferometric RM studies. 

We note that optical observations also provide additional constraints.  Besides constraining the SMBH mass, the HST data that detected the nuclear gas disk can be used to further infer the accretion rate associated with the gas disk. As described in Sect.~\ref{ssec:mass}, the mass estimated from the gas dynamics differs from those from the stellar dynamics and the EHT. The discrepancy could stem from inappropriate assumptions about the dynamics of the gas disk, as discussed in detail by \citet{Kormendy2013}. \citet{Jeter19} refined the gas-dynamical model by adopting a sub-Keplerian disk velocity together with a different viewing angle and demonstrated that the refined model can explain the larger BH mass.  By using the refined model with conservative estimates of the cold gas at 100\,pc scale by \citet{Tan08}, they estimated the mass accretion rate associated with the gas disk to be \textcolor{black}{$\sim$$2 \times 10^{-3}\,M_{\odot}\,{\rm yr}^{-1}$}.

\subsection{The event horizon scales}\label{ssec:accretion_horizon}

The accretion flow structure near event-horizon scales has become accessible through direct imaging with mm-VLBI. As described in Sect.~\ref{sec:mass}, the EHT observations at 1.3\,mm unveiled a ring-like structure in the core of M\,87~\citep{EHTC2019a, EHTC2019d, EHTC2019f}. By comparing the EHT images with an extensive library of GRMHD and GR radiative transfer (GRRT) simulated images, the observed ring-like structure was found to be in good agreement with the emission associated with the accretion flow \citep{EHTC2019e}. Further constraints on the properties of the 1.3\,mm ring-like structure were obtained based on the subsequent analysis of linear-polarization data \citep{EHTC2021a, EHTC2021b}. It was reported that the observed electric vector polarization angle (EVPA) distributions predominantly describe a nearly azimuthal pattern, suggesting that the poloidal magnetic field component is more dominant than the toroidal one at these scales. An extensive comparison of the EHT polarization images with a large GRMHD simulation library indicated that MAD is the favored state of the accretion flow, rather than SANE ones, suggesting the presence of dynamically important magnetic fields near SMBH. The recent further detection of circular-polarization signals in the EHT 2017 data additionally reinforces the MAD scenario~\citep{EHTC2023a}. 
Based on the GRMHD models that are in good agreement with the observations, the mass accretion rate at the EHT scales was estimated to be $\dot{M}\sim (3-20) \times 10^{-4}\,M_{\odot}\,{\rm yr}^{-1}$ \citep{EHTC2021b}.
Analysis of the EHT total-intensity and polarization images also provided horizon-scale estimates on other physical quantities such as electron density ($n_{\rm e} \sim 10^4 \text{--} 10^7 \,{\rm cm}^{-3}$), electron temperature ($T_{\rm e} \sim (1 \text{--} 12) \times 10^{10} \, \text{K}$), and magnetic field strength ($B \sim (1-30)$\,G).

\begin{figure}[ttt]
    \centering
    \includegraphics[width=\columnwidth]{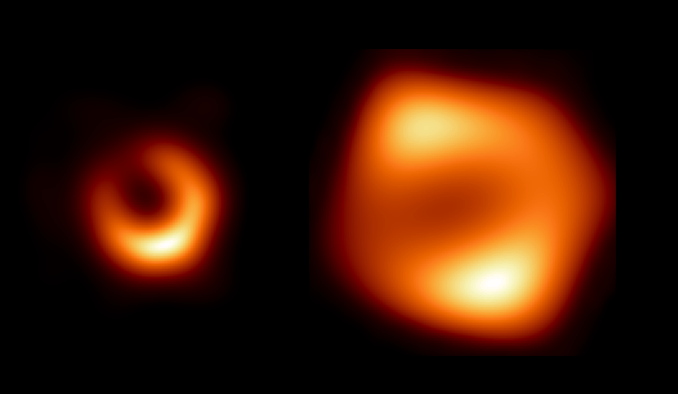}
    \caption{
    Spatially-resolved images of the central engine of M\,87 taken by EHT at 1.3\,mm (left) and GMVA at 3.5\,mm (right), placed side by side at the same scale. Both images were obtained from near-in-time observations made in April 2018. The EHT and GMVA images are adapted from the works of \citet{EHTC2024} and \citet{Lu2023}, respectively. The difference in the observed appearance is attributed to the opacity effects associated with the accretion flow surrounding the SMBH.
    }
    \label{fig:GMVA-AF}
\end{figure}

Following the EHT imaging of the BH shadow at 1.3\,mm, recent Global Millimeter VLBI Array (GMVA) observations connected to ALMA and the Greenland Telescope (GLT) have revealed another ring-like image in the M\,87 nucleus at a wavelength of 3.5\,mm~\citep[][see also Fig. \ref{fig:GMVA-AF}]{Lu2023}. The diameter of the 3.5\,mm ring is approximately 8.4 times the Schwarzschild radius, which is 50\% larger than the diameter of the 1.3\,mm ring. Additionally, the inner edge diameter of the 3.5\,mm ring is also larger than the diameter of the EHT ring. The larger and thicker ring size at 3.5\,mm is likely caused by the opacity effect of synchrotron emission at a lower frequency. A comparison of the GMVA image with GRMHD simulations in \citet{Lu2023} provides a strong indication that the observed ring-like structure at 3.5\,mm is dominated by the accretion flow.

\subsection{Evolution of $\dot{M}$ from Bondi to event horizon scales}\label{ssec:accretion_add}

As a summary of this section, in Fig.~\ref{fig:Mdot_sum} we show a plot compiling various estimates of $\dot{M}$ in the literature, measured at various scales.  Although the statistics are still very limited, a collection of these estimates reveals a substantial decrease in mass supply towards the innermost regions with respect to the Bondi accretion rate.  This is clear evidence that the classical ADAF is no longer valid as the model for the M\,87 accretion flows, while the suggested radial evolution of $\dot{M}$ is consistent with with ADIOS-like solutions.  

\begin{figure}[ttt]
    \centering
    \includegraphics[width=0.9\columnwidth]{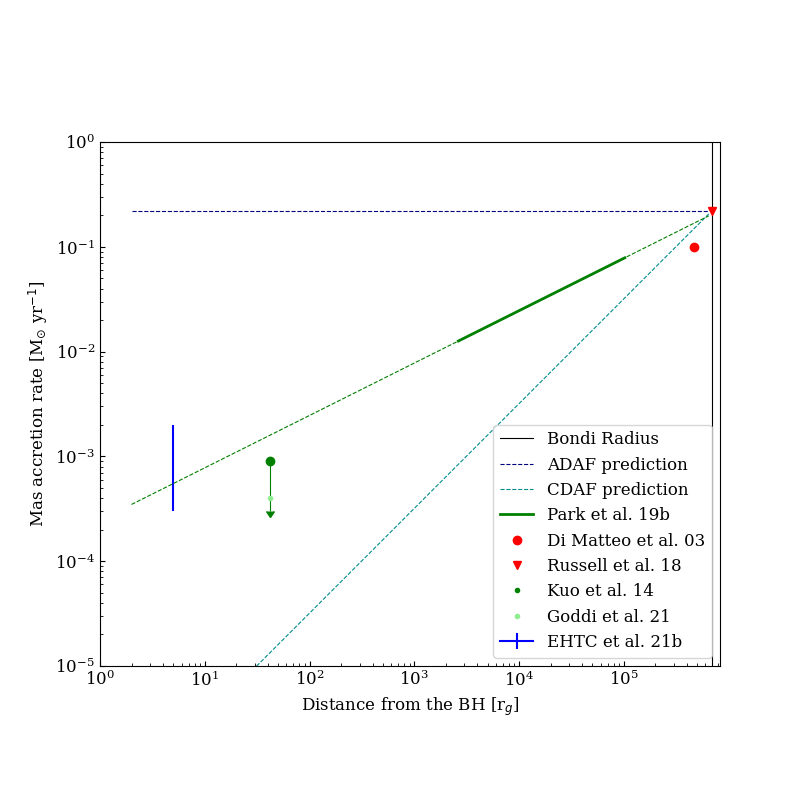}
    \caption{Radial evolution of M\,87 mass accretion rate, obtained by compiling various estimates at different distances from the core. Note that, although \citet{Goddi21} derived a mass accretion rate using the formula provided by \citet{Kuo14}, they remarked that RM in M\,87 might not provide an accurate estimate of $\dot{M}$ given the complexity and variability of the source structure.}
    \label{fig:Mdot_sum}
\end{figure}

\section{Relativistic jets}\label{sec:jet}
\subsection{Jet collimation}\label{ssec:collimation}

\subsubsection{Classical view}\label{sssec:jet-classical}
Narrowly collimated AGN jets extend well beyond their host galaxies on kpc scales. They sometimes propagate up to tens of kpc to Mpc scales, forming giant radio lobes as the end-point of jets where a huge amount of the energy, originally generated in the vicinity of SMBH, is deposited into the ISM or IGM \citep{blandford1990}. How can such a collimation of the highly relativistic plasma beam be sustained over many orders of magnitude in gravitational distance? On kpc scales, a number of AGN jets have been imaged with various instruments, commonly exhibiting conical structures with constant jet opening angles $\theta_{\rm j} \lesssim$ a few degrees. For example, the conical structure of the \M87 jet at $\gtrsim$ 1 arcsec was confirmed in 1970s--1980s with a photometric analysis \citep{deVaucouleurs1979} and VLA imaging \citep{owen1980, reid1982}.

On pc scales, according to the classical standard picture of AGN jets, the compact radio core (also referred to as the `VLBI core') seen at the apparent jet origin in a VLBI map has been widely believed as the throat of a diverging conical jet \citep{blandford1979} (see also Sect.~\ref{ssec:core}). Alternatively, \citet{daly1988} proposed that the VLBI core represents the first `re-collimation Mach disk–oblique shock' system, which is supposed to be located far from the BH ($\sim$$10^{4-6}\,r_{\rm S}$). 
Thus, the jet acceleration and collimation zone (ACZ) was supposed to exist in the upstream region of the mm-wave radio core, or may {\it not} exist \citep{marscher1985}.

\subsubsection{Jet collimation break}\label{sssec:JCB}
Based on early studies of the kpc-scale jet morphology \cite[e.g.,][]{reid1982} and the theory of MHD jets \citep{blandford1982, li1992}, the conical structure of the \M87 jet would suggest that the collimation of the jet \citep{junor1999}, as well as the jet bulk acceleration, seemed to terminate on pc scales. However, there was no plausible/coherent picture of the ACZ on AGN jets until mid 2010s (see also Sect.~\ref{sssec:acceleration}). The upper panel of Fig.~\ref{fig:ACZ} overviews the global geometry of the \M87 jet obtained by compiling the literature \cite[see the caption for references,][]{Nakamura18}. 

Multi-frequency VLBI observations by \citet[][]{asada2012} revealed that the global structure of the jet sheath is characterized by a semi-parabolic stream $z \propto R^{1.73}$ at $z \sim 400 - 4 \times 10^{5}\,r_{\rm g}$ (dashed line in Fig.~\ref{fig:ACZ} upper panel), while it changes into a conical stream $z \propto R^{0.96}$ (solid line in Fig.~\ref{fig:ACZ} upper panel) beyond the Bondi radius.  
This provides the first observational evidence of the jet collimation break (JCB) in AGN jets as a possible consequence of the gravitational interaction between the SMBH and host galaxy: the ACZ may be self-regulated under these co-evolving system. After the discovery of JCB in the M\,87 jet, subsequent investigations on many other jet sources also found similar jet geometrical transitions \citep[e.g.,][]{tseng2016, nakahara2018, nakahara2019, nakahara2020, Hada18, park2021b, Boccardi21, kovalev2020, okino2022}. 

Near the jet base of \M87, a parabola-like profile further continues down to a few tens of $r_{\rm g}$ from the BH~\citep{Nakamura2013, Hada13, Hada16, Kim18b}, while the recent GMVA+ALMA+GLT 86\,GHz image has detected a wider emission width than expected from the parabola closer to the BH~\citep{Lu2023}. This additional emission could be assocaited with the winds from RIAF supporting the initial jet collimation phase.

\begin{figure}[ttt!]
    \centering
    \includegraphics[width=\columnwidth]{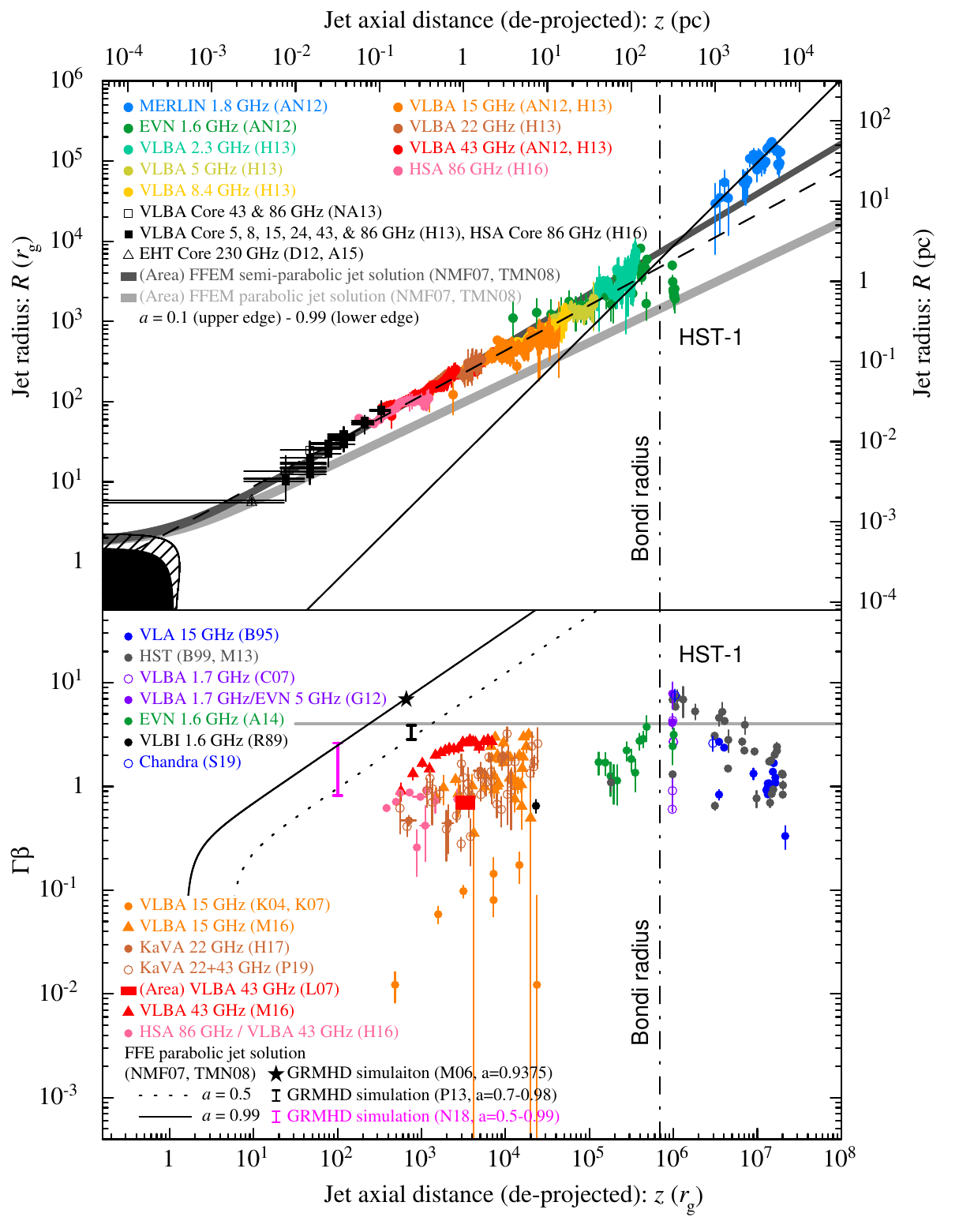}
    \caption{(Top): Distribution of the jet radius $R$ as a function of the M87 jet axial distance $z$ from SMBH (a de-projection with $\theta_{\rm view} = 14^{\circ}$ is adopted) in units of $r_{\rm g}$. (Bottom): Distribution of the four-velocity
    with additional VLBI data points \citep[KaVA 22 + 43 GHz from][]{Park2019a} as well as X-ray data points \citep[Chandra from][]{Snios2019}. Both panels are updated from \citet[][]{Nakamura18}.}
    \label{fig:ACZ}
\end{figure}

\subsubsection{Testing magnetically organized ACZ}\label{sssec:testmagacz}
For comparison with the observed data, the upper panel of Fig.~\ref{fig:ACZ} additionally overlays the outermost streamlines (anchored to the event horizon with the maximum colatitude angle of $\pi/2$) of the
semi-analytical, force-free electrodynamic jet model \citep{Narayan2007, Tchekhovskoy2008} with different BH spin parameters ($a=0.5-0.99$). The observed jet profile match well with the theoretical expectations of a semi-parabolic stream, although they deviate beyond the Bondi scale.

Also, high-resolution GRMHD simulations \citep{Nakamura18} exhibit a large Lorentz factor ($\Gamma$) sheath as a consequence of the so-called `magnetic nozzle' effect \citep[][]{camenzind1987, komissarov07, tchekhovskoy2009} along the funnel edge in the form of a parabola. This could be a theoretical analog to the `limb-brightened' structure seen in the \M87 jet. A detailed analytical study, using a steady-state axisymmetric force-free jet model, reproduces symmetrically limb-brightened radio images when the jet's magnetic field lines pass through a fast-spinning BH, while this is not the case for the magnetic field lines passing through a slowly-spinning BH or a Keplerian accretion disk \citep{takahashi2018}. Thus, the quasi-symmetric limb-brightend jet feature seen in \M87 would imply a spinning BH-driven jet.

\subsection{Jet dynamics}\label{ssec:dynamics}

\subsubsection{Radial evolution}\label{sssec:acceleration}
Compared to the well-established jet collimation profile, the dynamics of the \M87 jet remain largely controversial. The bottom panel of Fig.~\ref{fig:ACZ} summarizes the observed velocity distributions (converted to four-velocity $\Gamma \beta$) as a fuction of jet axial distance. On kpc scales, the proper motion of the jet is relatively well characterized~\citep[][see also Sect.~\ref{ssec:kpc}]{Biretta1999, Meyer2013, Snios2019}, showing an overall superluminal motion that decelerates from $\sim$6\,$c$ (corresponding to $\Gamma \beta \sim$7--9) at a deprojected distance of $\sim$200\,pc (HST-1; Sect.~\ref{ssec:hst-1}) to $\sim$0.5\,$c$ at distances greater than 1000\,pc. However, on pc-to-subpc scales, early VLBI monitoring programs conducted at monthly or yearly intervals often reported very slow speeds~\citep{Reid1989, Dodson2006, Kovalev2007, Ly2007}. These low speeds are inconsistent with the observed large jet-to-counterjet brightness ratio, implying that the sparse monitoring intervals are insufficient to accurately measure the inner jet motion.

A transition from subluminal to superluminal motion at intermediate scales between the jet base and HST-1 was initially reported by \citet{Asada2014} with the European VLBI Network (EVN) observations at 1.6\,GHz. Subsequently, multiple groups conducted more frequent monitoring programs of the \M87 inner jet using VLBA or East Asian VLBI Network (EAVN) at 22/43\,GHz~\citep{Walker2016, Mertens2016, Hada2017a, Walker2018, Park2019a}. These studies consistently detected fast motions (up to $\sim$2--3\,$c$) within a few 100\,$r_{\rm S}$ from the jet base, accounting for the large jet-to-counterjet ratio. They also found a trend of gradual acceleration with distance, culminating in maximum speed at HST-1. The observed acceleration regions appear to be coincident with those where the jet shape is parabolic, indicating a close connection between collimation and acceleration over a wide range of distances from the BH.

Nevertheless, these studies also claimed the presence of multiple velocity components even at the same radial distances from the BH (seen as large data scatters on pc scales in the bottom panel of Fig.~\ref{fig:ACZ}), implying that the true jet velocity fields are likely more complex. This may suggest that one speed is associated with a bulk flow while another traces a pattern or instability filament~\citep{Mertens2016}. Alternatively, the jet may contain velocity stratification where different jet speeds represent different layers within the jet~\citep{Park2019a}. Further accumulation of velocity measurements is needed to distinguish between these scenarios.

\subsubsection{Transverse evolution}\label{sssec:oscillation}

It has been relatively less known that the inner jet of \M87 exhibits morphological evolution perpendicular to the jet axis. This phenomenon was initially reported by \citet{Walker2016} and \citet{Walker2018}, who compiled a set of VLBA 43 GHz images spanning 17 years and observed a quasi-sinusoidal oscillation of the pc-scale jet with a period of approximately 8--10 years. Similar year-scale oscillations of the \M87 pc-scale jet were also reported at 15\,GHz~\citep{Britzen2017}. More recently, \citet{Cui23} have provided a significant update on the analysis of the jet transverse motion by compiling a large dataset of EAVN/VLBA 43/22 GHz data spanning 22 years from 2000 to 2022. They identified two cycles of oscillation in the jet position angle, implying a periodicity with a period of $\sim$11 years.

The origin of the observed long-term oscillation is subject to debate. \citet{Walker2018} interpreted the variation as a Kelvin-Helmholtz (KH) instability mode developed at the boundary between the jet and the external medium. 
Alternatively, \citet{Cui23} proposed that the observed periodic oscillation could be explained if the jet base is precessing, likely via a Lense-Thirring effect caused by a tilted accretion disk with respect to the SMBH spin.

Besides the year-scale variations, \citet{Ro2023a} recently reported that the \M87 jet exhibits transverse fluctuations on much shorter timescales (roughly one year or less). Although the exact origin of this type of oscillation remains unclear, it could be associated with the propagation of jet HD/MHD instability or perturbed mass accretion occurring in the magnetically-arrested disk.

\begin{figure}[ttt]
    \centering
    \includegraphics[width=\columnwidth]{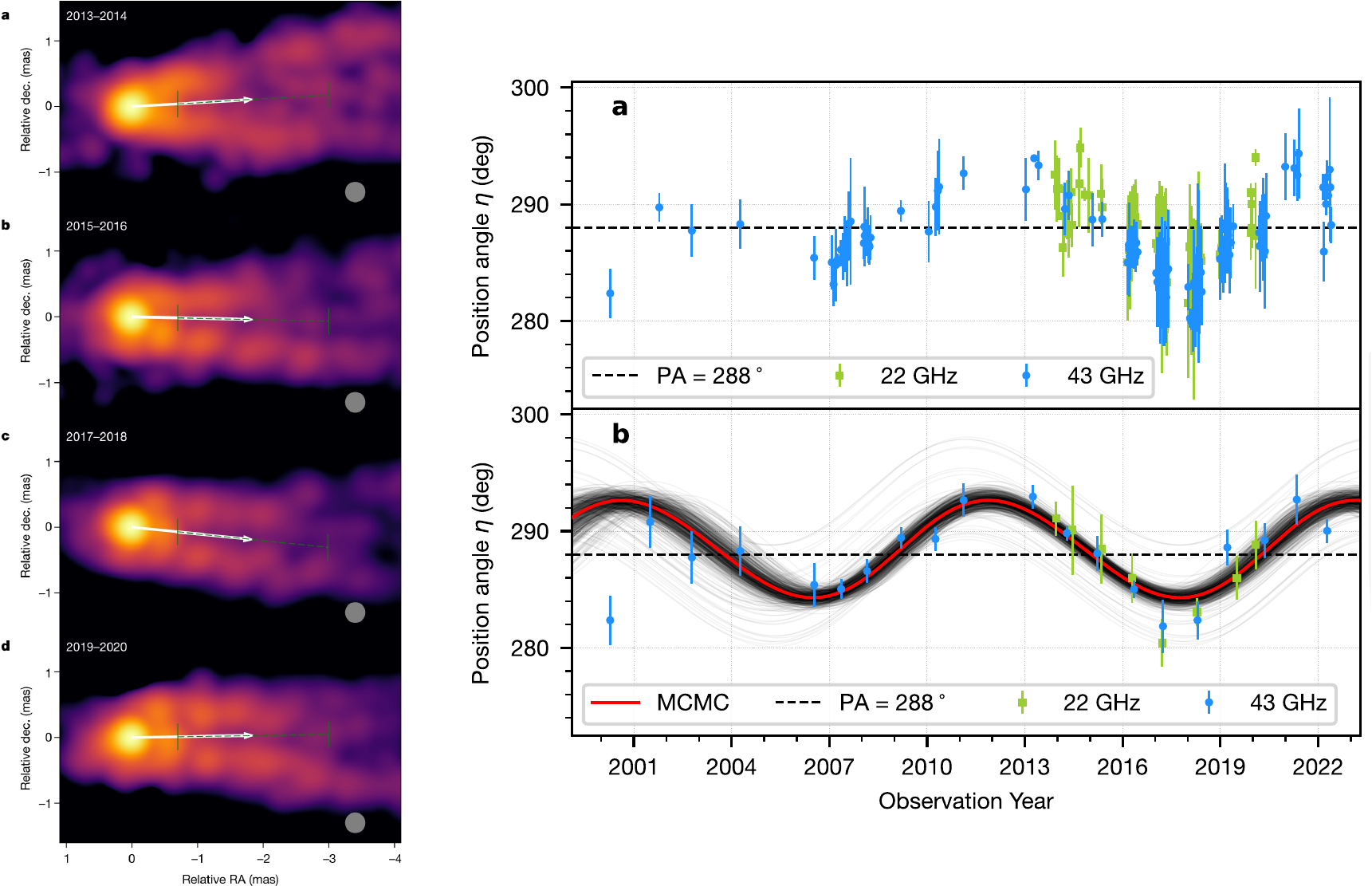}
    \caption{(Left) Long-term evolution of the M\,87 pc-scale jet monitored by VLBA/EAVN at 43\,GHz. (Right) Evolution of jet position angle. Images adapted from \citet{Cui23}.}
    \label{fig:enter-label}
\end{figure}

\subsection{Radio core}\label{ssec:core}
While the EHT 230\,GHz observations resolved the shadow of SMBH at the jet base, it is worth noting the nature of the `radio core' seen at the lower radio frequencies. The radio core, a bright compact feature at the apparent origin of an AGN jet seen in a VLBI image, is most likely either a synchrotron-self-absorbed opaque surface of the jet base at a given frequency~\citep[e.g.,][]{Konigl81, Lobanov98} or a standing shock feature~\citep[e.g.,][]{daly1988, Marscher08}. In the case of \M87, multi-frequency VLBA astrometric observations measured a core-shift between 2 and 88\,GHz with a frequency dependence of $\sim \nu^{-(0.9-1.0)}$~\citep{Hada11, Jiang21}, in agreement with the \M87 radio core at these frequencies being a photosphere of the jet base within a few tens $r_{\rm S}$ from the central BH. On the other hand, as introduced in Section~\,\ref{sec:accretion}, the recent GMVA+ALMA+GLT 86\,GHz observations have spatially resolved the radio core into a ring-like structure, with emission likely dominated by the inner part of accretion flows~\citep{Lu2023}. Thus at $\sim$3\,mm, the emission from both the accretion flows and the jet base could be blended when the radio core is imaged with a lower angular resolution. 

At cm wavelengths. a handful of ultra-high-resolution space-VLBI observations resolved the radio core region into highly complex morphology~\citep{Dodson2006, Asada2016, Kim23}. Especially at lower frequencies (1.6/5\,GHz), space-VLBI images revealed diffuse substructures surrounding the compact core perpendicular to the jet axis~\citep{Dodson2006, Asada2016}. The nature of this low-level emission has not yet been explored in detail, but it may hint at the presence of transverse stratification in the base of jet/outflows. 

\subsection{Counter jet}\label{ssec:cj}

The jet of \M87 is predominantly one-sided on both kpc and pc scales, indicating its strong Doppler-boosted nature. This suggests that the viewing angle of the jet is relatively small or/and the jet speed is highly relativistic at these scales. On parsec scales, hints of the counter jet were observed in some early VLBI images~\citep[e.g.,][]{Reid1989,Junor95}, but it did not receive much attention until \citet{Ly2004} claimed the first detection of a faint counter jet at the eastern side of the radio core, although the possibility of a calibration artifact was not entirely ruled out. 

The (sub)pc-scale counter jet of \M87 is now routinely observed in high-quality VLBI images at various frequencies (Fig.\ref{fig:cj}), and careful imaging tests by multiple teams demonstrate that this feature is very likely real~\citep[e.g.,][]{Kovalev2007, Ly2007, Kim2018, Walker2018, Nikonov23}. Interestingly, \citet{Kovalev2007} and \citet{Walker2018} reported that the counter jet was also edge-brightened similar to the main jet, and a similar structure was also identified in more recent study by \citet{Nikonov23} (Fig.~\ref{fig:cj}). The counter jet appears to extend $\gtrsim$3\,mas from the core at 15\,GHz, but looks even more extend if we take a look at the high-sensitivity VLBI image at 8\,GHz presented in \citet{Nikonov23}. The kinematics of the counter jet are much less constrained than that of the main jet, but independent measurements by different groups~\citep{Kovalev2007,Ly2007,Hada16,Mertens2016} consistently reported substantially slow apparent speeds ($\sim$0.01--0.17\,$c$).   

The jet-to-counter-jet brightness ratio depends on various factors such as observing frequency, core distance, possible variability, and whether the ratio is calculated using integrated fluxes or peak intensity. However, a typical range would be of the order of 10 at $\sim$1\,mas from the core. Combining this with proper motion measurement of the jet and counter jet, a conservative estimate of the jet viewing angle results in $\theta_{\rm view}$$\sim$13--40$^{\circ}$. A tighter constraint of $\theta_{\rm view}$$\sim$14--20$^{\circ}$ is obtained by considering more detailed jet velocity field~\citep{Mertens2016}, consistent with the angle suggested from the HST-1 superluminal motion~(Sect.~\ref{ssec:hst-1}).

\begin{figure}[ttt]
    \centering
    \includegraphics[width=\columnwidth]{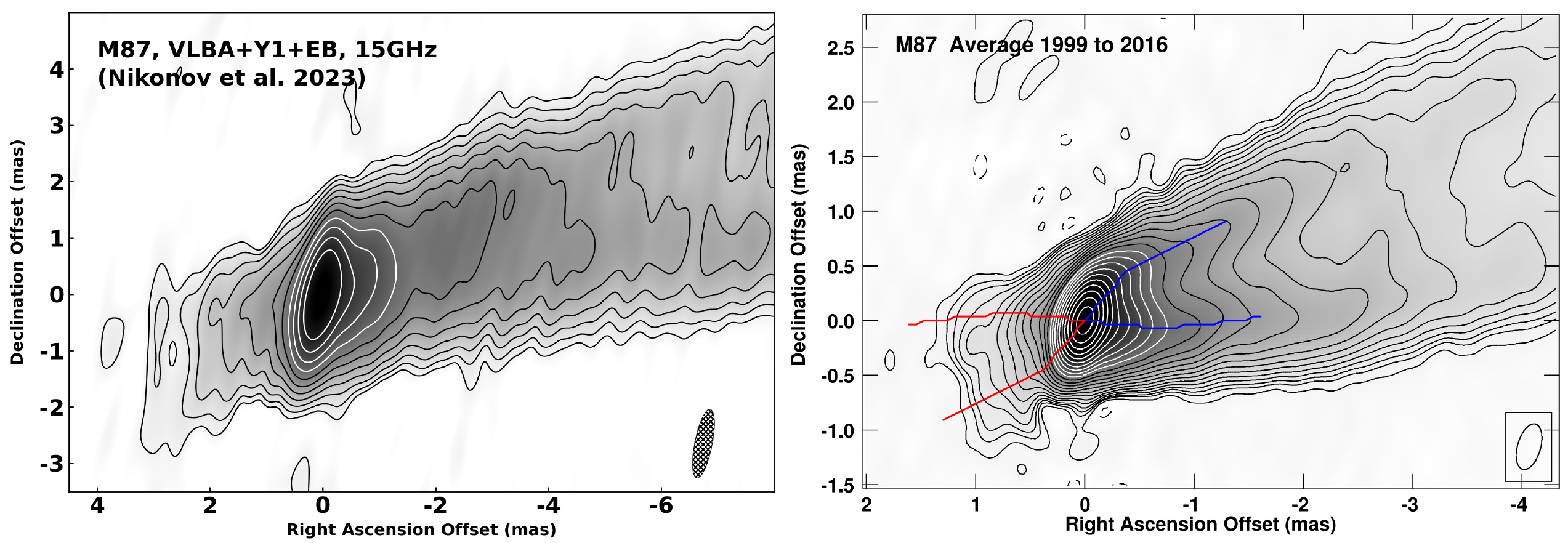}
    \caption{(Left) counter jet of \M87 imaged with VLBA+Y1+Effelsberg at 15\,GHz. The image was generated by using a data provided by \citet{Nikonov23}. (Right) M\,87 counter jet image obtained with VLBA at 43\,GHz. The image was reproduced with permission from \citet{Walker2018}, copyright by the AAS.}
    \label{fig:cj}
\end{figure}

\subsection{Polarization}\label{ssec:polarization}

Since the first identification of polarized emission from \M87 by \citet{baade1956}, the polarimetric properties of this jet have been extensively investigated across various scales and wavelengths. Polarimetry serves as a powerful tool for probing the structure of magnetic field, the internal jet structure as well as interactions between the jet and its surrounding environment. 

On kpc scales, the polarization of the \M87 jet has been studied in great detail since 1980s using VLA in cm radio bands~\citep[e.g.,][]{owen1989,Owen1990} and later also using HST in optical bands and ALMA in mm bands~\citep[e.g.,][]{Capetti1997,Perlman1999,Avachat2016,Goddi21}. Polarized emission is detected throughout jet regions on kpc scales, with enhanced signals predominantly associated with the bright knot regions. While the observed magnetic field position angles (MFPAs; formally $90^{\circ}$ rotated from the measured EVPAs) on kpc scales are largely parallel to the jet axis in both radio/optical bands, only the optical MFPA vectors tend to become perpendicular to the jet at the upsteam end of some bright knots, suggesting a multi-layered structure in the jet~\citep{Perlman1999}. 
More recently, analyzing a wideband (4--18\,GHz) Jansky VLA full-polarimetric data, \citet{Pasetto2021} revealed extremely detailed radio polarization and Faraday RM distributions that support the presence of a helical magnetic field on kpc scales. See also Sect.~\ref{ssec:kpc} for a review on the kpc-scale jet.  

In contrast to the rich polarization structures seen at kpc and horizon scales (see Sect.~\ref{ssec:accretion_horizon}), still much less is known about the polarimetric properties on intermediate ($\sim$0.1--100\,pc) scales, as the pc-scale jet of \M87 is largely unpolarized at radio wavelengths. Pioneering studies were conducted by \citet{Junor2001} and \citet{Zavala2002} using VLBA at cm wavelengths (5--15\,GHz), where they detected patchy polarization features in the jet portion around $\sim$20\,mas from the core with considerably large RM magnitudes (thousands to $\sim$$10^{4}$\,rad\,m$^{-2}$). Weak patchy polarization features were also detected closer to the core at 86\,GHz~\citep{Hada16}. A significant advance was made by \citet{Park2019b}, where they revealed more detailed RM distributions along the pc-scale jet by revisiting multi-frequency polarimetric VLBA data at 2--8\,GHz. They found that the RM magnitude gradually decreases with distance from one parsec scales to just before HST-1. The observed slope of RM distributions was reproduced with the scenario that the Faraday screen is the winds from the hot accretion flows, which may be relevant to the observed collimation and acceleration of the M87 jet on the same scales. More recently, \citet{Nikonov23} have reported the presence of an RM gradient also in the direction perpendicular to the jet based on their high-sensitivity VLBI images at 8 and 15\,GHz.  

Between the horizon scales and parsec scales, the polarization structure of the radio core at subpc scales remains controversial. \citet{Walker2018} and \citet{Kravchenko2020} reported complex EVPA patterns that appeared to be wrapped around the 43\,GHz core. However, a later analysis by \citet{Park2021a} with an improved calibration method did not find such a complex feature but detected a simple compact polarization component with its peak coinciding with the total intensity peak. To obtain a definitive answer and bridge the magnetic field structure between horizon and parsec scales, higher sensitivity VLBI observations at 43/86\,GHz combined with a refined polarization calibration technique would be required.

\begin{figure}[ttt]
    \centering
    \includegraphics[width=\columnwidth]{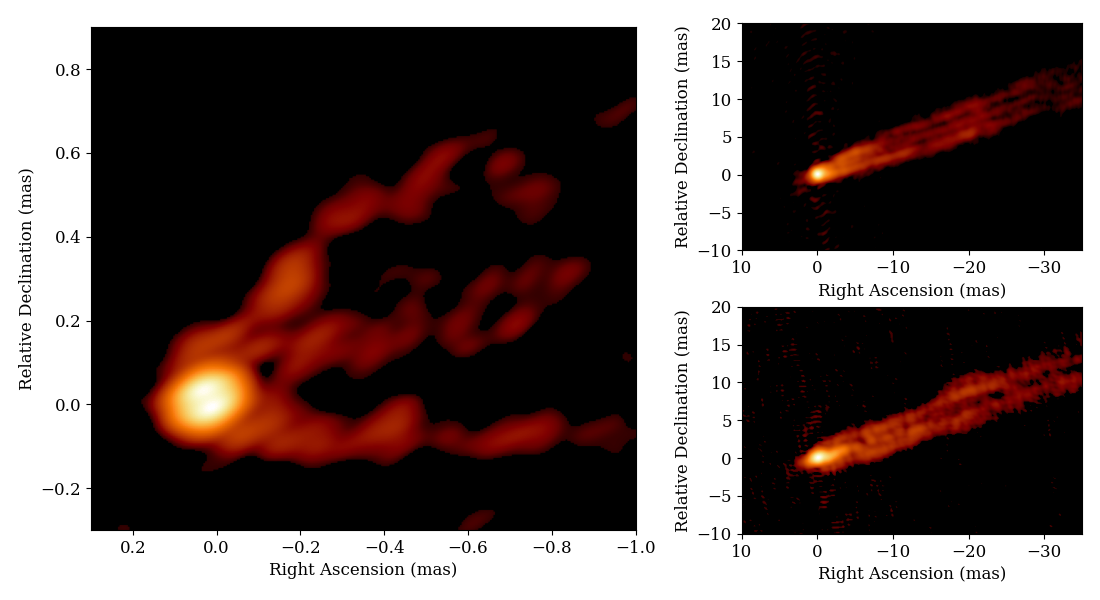}
    \caption{Triple-ridge structure of the M\,87 jet observed at various frequencies and core distances. (Left) GMVA+ALMA+GLT image at 86\,GHz~\citep{Lu2023}. (Right-top) High-sensitivity Array image at 15\,GHz~\citep{Hada2017b}. (Right-bottom) VLBI Space Observatory Programee (VSOP) image at 5\,GHz~\citep{Asada2016}. 
    }
    \label{fig:spine}
\end{figure}

\subsection{Spine-sheath structure}\label{ssec:spine-sheath}
As mentioned in Sect.~\ref{ssec:collimation}, the pc-scale jet of \M87 is widely known to show a limb-brightened structure with a parabolic shape. However, when imaged at a higher resolution or/and sensitivity, another streamline appears to emerge in the middle of the jet (Fig.~\ref{fig:spine}). Such a `spine' component was clearly detected in high-quality VLBI images at cm wavelengths~\citep{Asada2016,Hada2017b,Tazaki23,Nikonov23} and also in stacked VLBI images at 43/86\,GHz~\citep{Walker2018,Kim2018}. The central spine emission appears to be substantially narrower than the whole jet width.
Recently, a new GMVA 3\,mm M87 image connected to ALMA and GLT has clearly detected a triple-ridge jet structure even closer to the core~\citep{Lu2023}. These images obtained at various scales/frequencies indicate that such a `spine-sheath' structure of this jet is maintained over a wide range of distances from the BH vicinity to further out.

The nature of the observed triple-ridge profile remains a matter of debate, with some potential scenarios proposed to explain the origin. One possibility is that the central ridge is actually part of the outer sheath associated with the same layer as the northern/southern limbs, while the true spine emission near the jet axis is less dominant due to either a low emissivity or deboosting~\citep{Mertens2016, Walker2018}. 
solely reproduced by a BH-powered jet~\citep{Asada2016, Orihara2019}, with different streamlines or beaming factors. 
Another possibility is that the central ridge is associated with the true spine within the jet interior, which may be anchored in the central rotating BH~\citep{Asada2016, Sobyanin17, Ogihara2019}.

\subsection{HST-1}\label{ssec:hst-1}
At a deprojected distance of $\sim$200\,pc from the nucleus, the \M87 jet harbors a notable feature known as ‘HST-1’. This feature was initially discovered with HST by \citet{Biretta1999}, and has been attracting special attention from the high-energy astrophysical community for several reasons.

First, HST-1 contains highly superluminal components of $\sim$(4--6)\,$c$ that are observed across radio, optical and X-ray bands~\citep[e.g.,][]{Biretta1999,Cheung07,giroletti12, Snios2019}. The observed apparent speeds appear to be the highest in the entire velocity field of the M87 jet~\citep[][see also Fig.~\ref{fig:ACZ} bottom]{Meyer2013, Asada2014}.  This suggests that HST-1 represents a maximally Doppler-boosted region in the \M87 jet, and also tightly constrain the jet viewing angle to be $\theta_{\rm view}\lesssim 19^{\circ}$. Second, as elaborated further in Sect.~\ref{ssec:mwl}, HST-1 experienced a notable MWL outburst around 2005~\citep{harris06}, followed by the ejection of superluminal components~\citep[][Fig.~\ref{fig:hst1} top]{Cheung07}. 
Third, HST-1 is located in the transition region of the jet geometry, where the jet shape changes from parabolic to conical. Additionally, there is a contraction of the jet cross section at HST-1 itself~\citep{AN2012}, accompanied by highly enhanced polarization and RM~\citep{Park2019b}. The combination of these observational characteristics has led to an intriguing hypothesis that HST-1 represents a recollimation shock region at the end of ACZ in the \M87 jet, releasing significant energy in the form of high-energy flares. HST-1 could therefore be a possible counterpart of the unresolved core of distant blazars. 

\begin{figure}[ht]
    \centering
    \includegraphics[width=0.8\columnwidth]{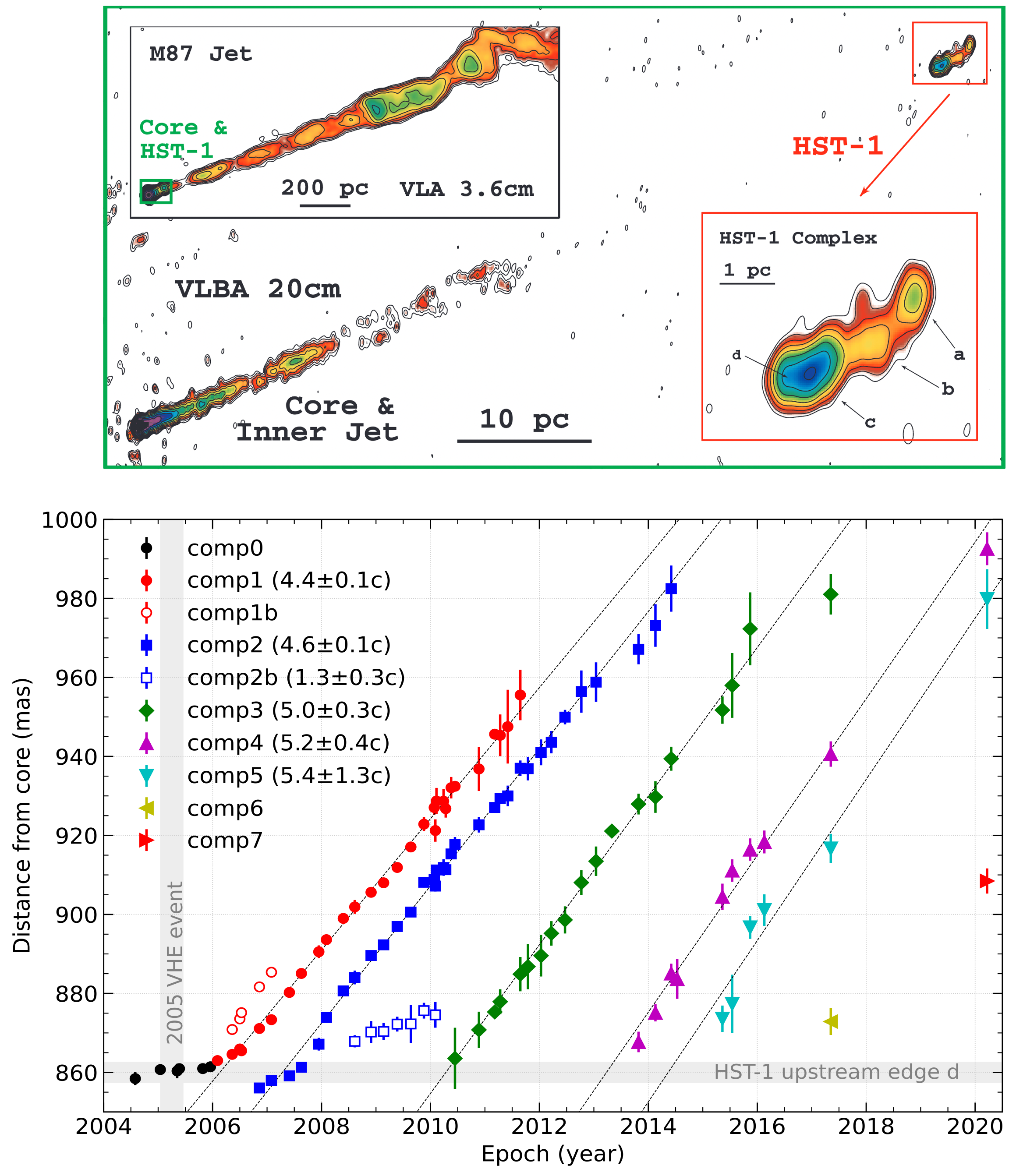}
    \caption{(Top) radio images of HST-1. The image was reproduced with permission from \citet{Cheung07}, copyright by the AAS. (Bottom) time evolution of radial positions of HST-1 subcomponents from 2004 to 2020. The plot is an update from \citet{giroletti12} and \citet{Hada2014a}.} 
    \label{fig:hst1}
\end{figure}

When viewed at the highest resolution, HST-1 exhibits intricate internal structure and time evolution~\citep[e.g.,][]{Chang2010}. It consists of multiple substructures with different velocities and trajectories. Long-term VLBI monitoring observations of HST-1 detected repeated ejection of new superluminal components from the upstream edge of the HST-1 complex~\citep[`HST-1d';][see also an updated plot in Fig.~\ref{fig:hst1}]{Cheung07, giroletti12, Hada2014a}, suggesting that the upstream edge is where particles passing through are reenergized. After the ejection, while all the components are ejected from a similar location, their subsequent trajectories are quite different from one another, with most components following curved paths. This implies that the actual trajectories of these components are three-dimensional. The presence of helical motion in HST-1 is suggested by \citet{Chen2011}, where they detected a progressive rotation of HST-1's EVPA using VLA data around the 2005 event. Unfortunately, the brightness of HST-1 has been continuously decreasing since the 2005 event across all wavelengths, rendering it very faint and challenging to observe as of 2024. Reenergization of HST-1 is eagarly awaited.

\subsection{Kpc-scale jet}\label{ssec:kpc}
\subsubsection{Spatially decomposed ideal laboratory}\label{sssec:kpc-1}
The \M87 jet on kpc scales has been well studied across a wide range of wavelengths from radio to X-rays over four decades. A conical shape with an opening angle of $\sim 6^{\circ}$ \citep{deVaucouleurs1979, owen1980, owen1989, reid1982} starts from the innermost bright knot G, lying about 1 arcsec from the nucleus in VLA observations \citep{owen1989}.
This region was later resolved by the HST into a structured complex known as HST-1 as introduced in the previous section. The structure of the jet downstream of HST-1 (1--18 arcsec or 0.1--1.5\,kpc in projected distance) can be characterized by trailing clumps or knots of bright gas (HST-1 to C: Fig. \ref{fig:kpc}) with an apparent deceleration to subluminal speed \cite[around 6\,$c$ to $\sim$0.5\,$c$][]{biretta1995, Biretta1999} and filamentary structures \cite[`wiggles/kinks'][]{owen1989, sparks1996}.

\begin{figure}[ttt]
    \centering
    \includegraphics[width=\columnwidth]{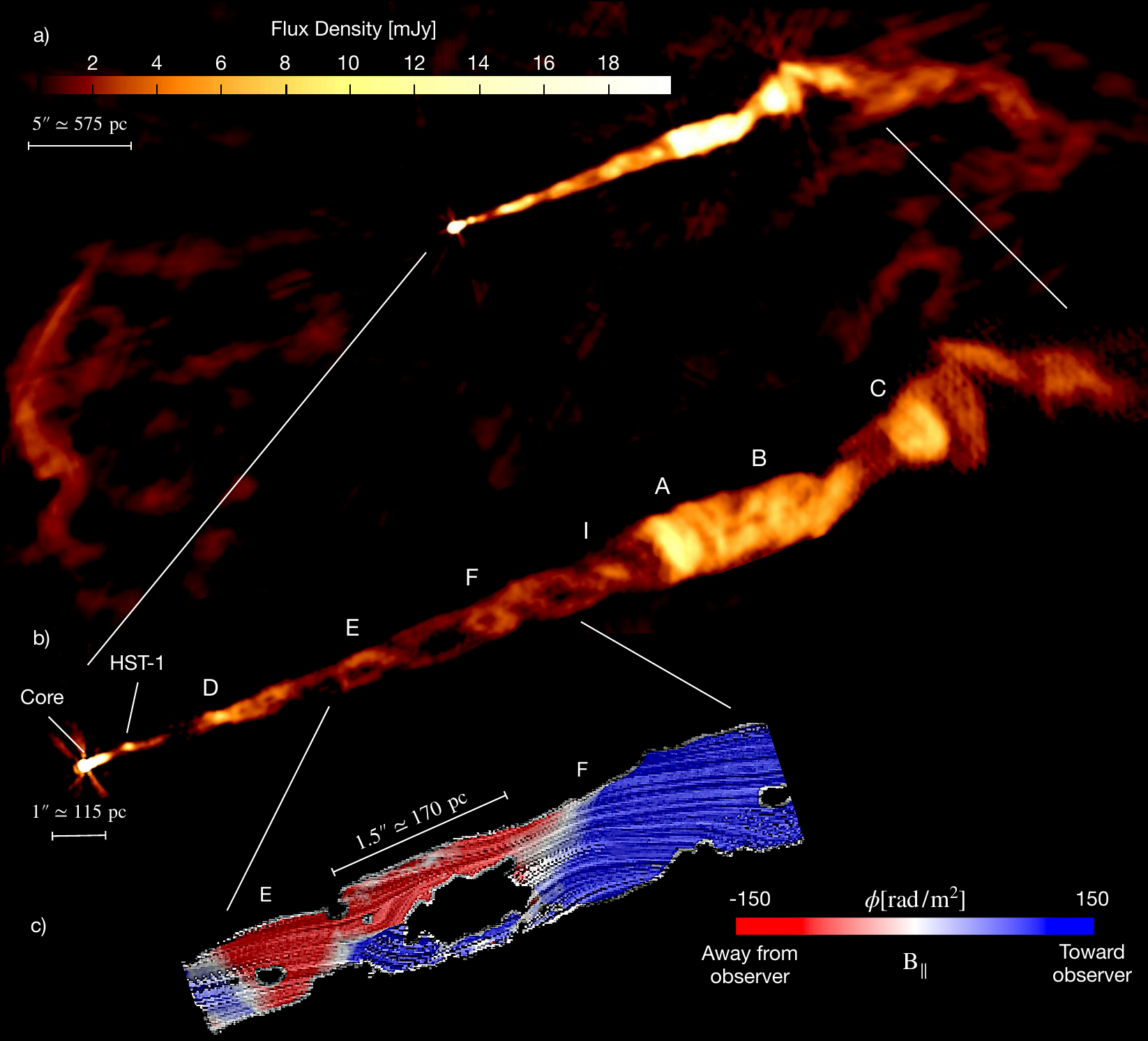}
    \caption{Jansky VLA total-intensity and RM images of M\,87 kpc-scale jet, obtained by combining wideband (4--18\,GHz) radio data. Figure produced by courtesy of Alice Pasetto.}
    \label{fig:kpc}
\end{figure}

The high overpressure in the synchrotron gas and the highly polarized (in both the knot: 40--60\,\% and interknot: 20--40\,\%) helical filaments \citep{owen1989, Perlman1999} indicate the existence of underlying ordered magnetic fields with a three-dimensional helix seen in projection; magnetic fields, therefore, appear to play a role in determining the \M87 jet structure even on large scales. While the projected MFPA vectors in the kpc-scale jet typically lie along the jet, the vectors at the brightest edges of the knots HST-1, D (especially in the optical band), A, and C becomes  perpendicular to the jet, indicating the presence of longitudinal compression by a shock front and/or a tightly wound magnetic helix \citep{owen1989, Perlman1999}.  Detailed broadband (from radio through optical to X-ray) spectral shape of the knots (HST-1 to C) in the \M87 jet favors the scenario in which synchrotron emission dominates the radiation and in-situ particle acceleration (by the first-order Fermi process) almost certainly occurs in the large scale \M87 jet \citep[both within knots and outside them;][]{Perlman05}.

Under the assumption of the minimum energy condition, the knots themselves appear to be significantly overpressured \citep{owen1989} with respect to the ambient thermal gas \citep{young02}, but the interknot regions do not \citep{sparks1996}. In order to maintain a conical streamline of the adiabatic jet within a uniform ambient gas (an isothermal King profile with a core radius $r_{\rm c} \simeq 18$ arcsec), the fields may have to be much stronger and more highly ordered than a weak and tangled field at the equipartition level ($\sim$ a few of 100\,$\mu$G). Magnetic fields, therefore, appear to play a crucial role in determining the structure of the \M87 jet even on larger scales beyond 100 pc.

\subsubsection{Origin of the comprehensive structure}\label{sssec:kpc-2}
A MWL study of the \M87 jet during 2002--2009 examined by \citet{Avachat2016} reveals some differences in polarimetric and spectral features compared to the earlier result \citep{Perlman1999}; a number of regions in the jet have a helical morphology as inferred from the the MFPA distributions. The presence of a systematic helical magnetic field in the particle-dominated jet, presumably maintained by KH instabilities \citep{lobanov2003, hardee2011}, has recently been reported by utilizing Jansky VLA broadband full-polarization radio data \citep[Fig. \ref{fig:kpc},][]{Pasetto2021}; a clear `double-helix' morphology of the jet together with systematic transverse RM gradients is resolved in the scales of $\sim$0.3--1\,kpc in projection.

As shown in Fig. \ref{fig:kpc}, the trails of the bright emission structure (`knots') are prominent. The innermost region of this trend, the HST-1 complex, appears to be an ignition point. Knots are interpreted as internal shocks in a collimated plasma beam \citep{rees1978}. Of particular interest is the brightest emission structure A–B–C. Knots A and C have certain similarities \citep{owen1989, Perlman1999}; (1) bright transverse linear features (normal to the jet axis) indicative of a shock front \citep{biretta1983}, and (2) the dominance of transverse magnetic field suggesting ordered helical magnetic components. Visible side-to-side oscillation is also observed between these knots and MFPA vectors appear to follow the fluctuating jet axis in this part (including knot B). 

A sudden enhancement of emission at knot A, at the upstream edge of that knot, indicates a reverse shock and a rapid drop in emission at the downstream edge of knot C, suggesting that it is the corresponding forward shock \citep[][]{harris2006b}. Particle acceleration is associated with both the forward and reverse modes. Sudden changes in MFPA vector orientation strongly imply the existence of MHD fast/slow-mode waves; the transverse component of the magnetic field $B_{\perp}$ increases across a fast-mode front, while $B_{\perp}$ decreases across a slow-mode front. 
The brightest emission structures A-B-C could be quad-relativistic MHD shocks (forward fast/slow and reverse slow/fast), preceding knots from the HST-1 complex; as is shown in Fig. \ref{fig:hst1}, an ejection of the paired super/sub-luminmal components (comp 1/1b in Fig.~\ref{fig:hst1}) from HST-1d, the upsream of the HST-1 complex, is reproduced in relativistic MHD simulations  \citep{nakamura2010}. 

\begin{figure}[ttt]
    \centering
    \includegraphics[angle=-90, width=1.0\columnwidth]{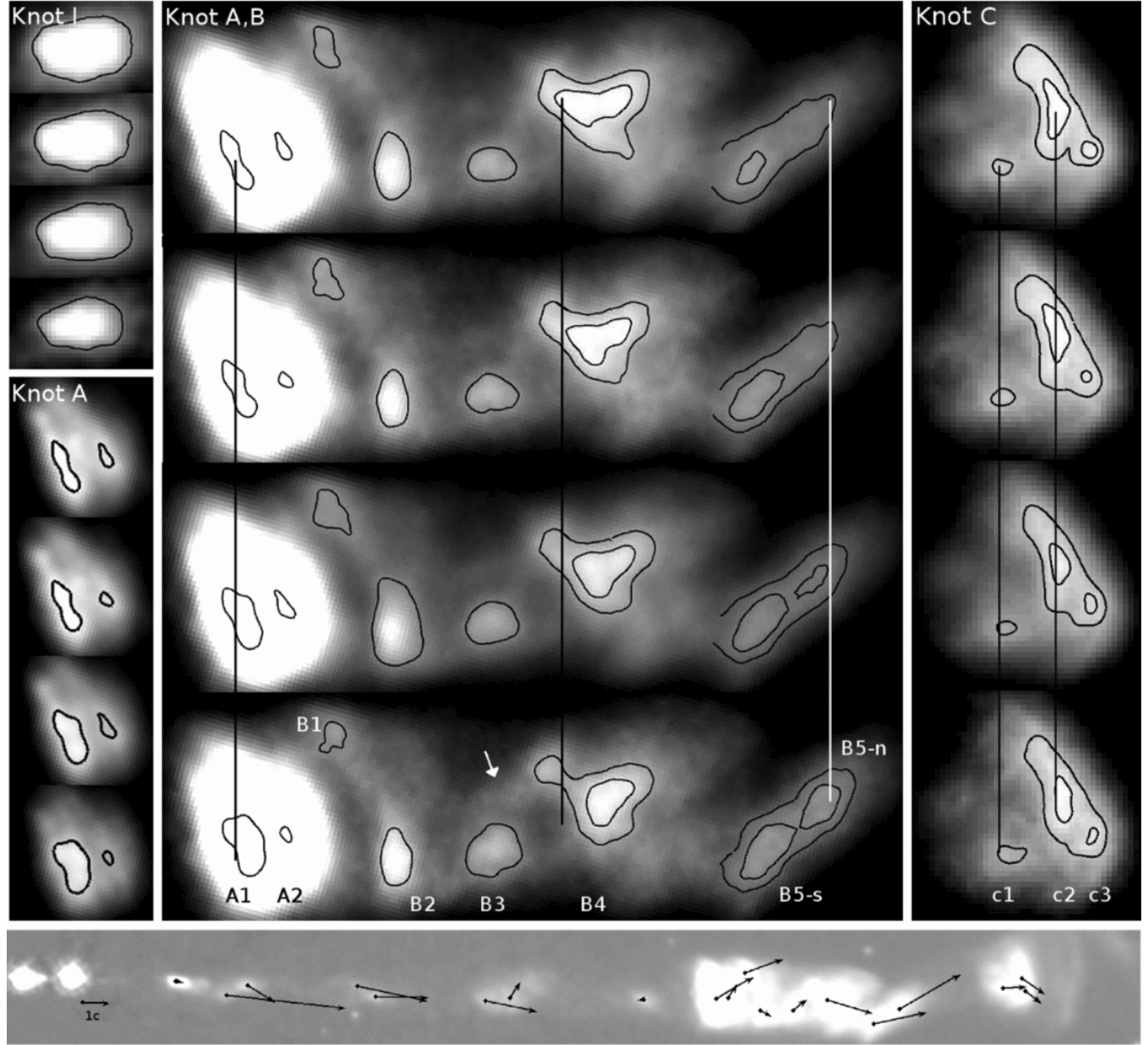}
    \caption{Motions and velocity vectors of the knots in the kpc-scale jet obtained with HST. Image reproduced with permission from \citet{Meyer2013}, copyright by the AAS.}
    \label{fig:kpc-hst}
\end{figure}

Similar patterns of trailing bright knots, together with the systematic orientation of the projected MFPA vectors, may also be accompanied by such `zigzag' patterns of the transverse velocity field (a flattened view of a helical motion) in HST observations \citep[][ Fig.~\ref{fig:kpc-hst}]{Meyer2013, nakamura2014}. Such a helical distortion could be seen as a consequence of growing kink mode ($m=1$) of the current-driven (CD) instability, triggered by the slow mode MHD shocks \citep[e.g.,][]{nakamura2004}. Readers may refer to some recent results of highly magnetized (the initial magnetization $\sigma \simeq 25$), relativistic MHD jet simulations \citep{barniol2017}; the kink instability is triggered when the jet passes through the recollimation region \citep{bromberg2016}. Supposing that this can apply to the \M87 jet, the HST-1 complex is responsible for the recurrent ejection of super/sub-luminal components and  the resultant `kinked segment' (unwinding of the magnetic coils) advect with the bulk flow \cite[{\rm e.g.}, knots D, E, F, and A-B-C:][]{Biretta1999, Meyer2013}.

We find an older, but suggestive simulation result of a highly magnetized, overpressured jet sheath against the uniform ISM environment \citep{clarke1986}, which may be applicable to a conical expansion in the M\,87 jet. In a system like M\,87, no cocoon surrounds the `naked' jet and the post-shocked jet material is confined by the Lorentz force to form a `nose cone' that leads the Mack disk \citep{burns1991}. In general, the surface-driven KH instability may be visible to the jet exterior, but it may be in conflict with the naked jet structure in M87. The growth of helical KH mode is suppressed even in a weakly magnetized jet with the axial magnetic field \citep{mizuno2007}.}

To address this issue (whether KH or CD instability dominates in organizing the kpc-scale jet in M\,87), we should consider the fraction of the dominant energy flux in jets. In magnetically driven jets, one of the standard models, the dominant Poynting flux near the BH (see also Sect. \ref{ssec:key-lj}) is converted into kinetic energy flux downstream, allowing the jet to achieve relativistic speeds. A transition from magnetically dominated to kinetically dominated states in jets would of course be expected further downstream (e.g. pc to kpc scales). The efficiency of this conversion (from Poynting to kinetic energy flux) depends on non-uniform lateral collimation or divergence of the poloidal magnetic filed lines inside the jet (magnetic nozzle effect; see also Sect. \ref{sssec:testmagacz}). As a consequence, a magnetically dominated jet may exist even on kpc scales (particularly in low-power FR-I sources like M\,87) even after the jet structural transition \citep[from parabolic to conical; e.g.,][]{barniol2017}. Further observational investigation would be encouraged.

\subsection{Broadband/MWL properties of (sub)pc-scale jet
}\label{ssec:mwl}

\subsubsection{Discovery of fast variability of VHE $\gamma$-ray emission}\label{sssec:vhe}

Until mid-2000, the only extragalactic
objects known to emit TeV $\gamma$-ray radiation
were blazars.
An early observational study with
High Energy Gamma Ray Astronomy (HEGRA) collaboration
reported weak evidence for TeV $\gamma$-ray emission from \M87 in the data during 1998 and 1999 \citep[]{Aharonian03}.
The first clear detection of fast variations of TeV $\gamma$-ray flux in \M87 was reported 
in 2005 by \citet{Aharonian06} using High Energy Stereoscopic System (H.E.S.S.). Similar phenomena were later confirmed by
Very Energetic Radiation Imaging Telescope Array System (VERITAS) and Major Atmospheric Gamma-ray Imaging Cherenkov Telescope (MAGIC)
\citep[]{Acciari08, Albert08}.
The variation time scales of $\Delta t \sim 2$~days
indicate that the characteristic 
size of the TeV $\gamma$-ray emission region is given by 
$R_{\rm TeV} \approx 5 r_{\rm S}\delta 
\left(\frac{\Delta t}{2~{\rm days}}\right)$,
where $\delta$ is the Doppler factor of the emission region
\citep[]{Aharonian06}.
While a majority of AGNs detected in VHE $\gamma$-rays are strongly beamed blazars, \M87 is weakly or at most moderately beaming.
Hence, the compactness of 
$R_{\rm TeV}$ likely indicates that TeV $\gamma$-rays are
produced in the vicinity of the BH.
The discovery of TeV $\gamma$-ray emission 
stimulated subsequent efforts of coordinated MWL observations of this source.

\subsubsection{The giant X-ray flare at HST-1 peaked in 2005}\label{sssec:xray}

During an intensive Chandra X-ray monitoring program of this source starting from early 2000s, a giant X-ray flare was detected from HST-1, whose X-ray flux peaked in 2005~\citep[]{harris03,harris06}. 
The X-ray intensity of HST-1 increased by more than a factor of 50 from 2000 to 2005. 
Year-scale light curves of HST-1 in optical and UV bands also 
showed the rise and decay synchronous with the X-ray flare \citep[]{Perlman11, Madrid09}.
Using the observed decay time scale 
for the X-ray flare,
the magnetic field strength at HST-1 can be 
estimated as 
$B=0.5~\delta^{-1/3}
    \left(\frac{\Delta t}{1~{\rm year}}\right)^{-2/3}
    \left(\frac{E}{1~{\rm keV}}\right)^{-1/3}
    {\rm mG}$,
where a synchrotron cooling 
is considered \citep[]{harris09}. 
Indeed, superluminal motions at HST-1 were 
clearly detected 
from the VLBI monitoring observations of the HST-1 components 
\citep[]{Cheung07,giroletti12}.
While the light curves of HST-1 in X-ray, UV, optical and VLA 2\,cm bands reached a maximum in 2005, 
the light curves of the nucleus in these bands showed no obvious correlation with the TeV flare in 2005. 
Therefore, the TeV emission in 2005 was suggested to originate in HST-1 and led to the hypothesis that HST-1 represents a blazar-like feature in distant jet sources \citep[]{haris08}.

\subsubsection{Location of VHE flares: nucleus vs HST-1}\label{sssec:vhe}
 
\M87 has been established as a VHE $\gamma$-ray emitter after the 2005 TeV event \citep[]{Aharonian06}. 
Since then,
exploring the origin of VHE $\gamma$-ray emitting region and its emission mechanism has been one of the major topics in the study of M\,87.
In \citet[]{Abramowski12},
long-term MWL light curves during 2001--2011 are summarized,  
where enhanced VHE activities were
detected in 2005, 2008, and 2010 by H.E.S.S.,
MAGIC and VERITAS.
This shed a new light on
the location of the VHE $\gamma$-ray emitting region in \M87.
    
In this context, near-in-time high-resolution VLBI monitoring observations coordinated with the $\gamma$-ray facilities have played a key role in localizing the site of high-energy events. 
To summarize, the VLBI data suggest that the $\gamma$-ray events that have occurred since 2008
have always originated from the nucleus.

\begin{itemize}
    \item
The H.E.S.S. Collaboration reported a higher $\gamma$-ray flux in 2005 than in the adjacent years~\citep{Aharonian06}. Owing to the observed rapid variability, they argued that the likely origin of the TeV emission was the nucleus (i.e. close to the SMBH).
On the other hand, \citet[]{Cheung07}
argued that the TeV emission originates from HST-1 that contains superluminal components rather than from the nucleus.

  \item

The MWL observing campaign in 2008 revealed a period of  
strong VHE $\gamma$-ray flares accompanied 
by a significant increase in the radio flux from the VLBA 43\,GHz core. 
This implies that VHE-emitting particles are produced in the vicinity of the central BH \citep[][and references therein]{Acciari09}.
As the flare progressed, the brightened region extended about 0.77\,mas from the core, resulting in the emergence of new components at an apparent speed of $\sim$1.1\,$c$.

    \item 
During the VHE event in 2010 April, VLBA observations (as well as Chandra) confirmed a quiescent state of HST-1. 
On the other hand, the VLBA radio core at 22/43\,GHz near in time to the VHE peak showed a slightly ($\sim$10\%) higher flux level than that of before/after the event \citep[]{hada12}.
The size of the 43~GHz core was estimated to be 17\,$r_{\rm S}$, 
which was comparable to the suggested size of the VHE emitting region. 
These results tend to favor the scenario that the VHE $\gamma$-ray 
flare in 2010 April is associated with the radio core.

    \item 
 During 2012 February--March, an elevated level of the \M87 VHE $\gamma$-ray flux was reported \citep{Beilicke12}\footnote{
Note, however, that a recent report by \citet{MAGIC20} indicates that the differential flux between 2012 and 2015 is at a similar level to that during the low-emission states reported by MAGIC between 2005 and 2007.}. 
 A densely-sampled radio monitoring program with the VLBI Exploration for Radio Astrometry (VERA) at 22/43\,GHz detected a significant flux increase from the radio core coincident with the VHE activity~\citep{hada14}.
 Meanwhile, EVN observations confirmed a quiescent state of HST-1 in terms of its flux density and mas-scale structure. The VLBI results, combined with a coincident proto-EHT 230\,GHz data~\citep{akiyama15}, suggested that the active event in 2012 originates in the jet base within $\sim$56\,$r_{\rm S}$ near the BH.

\end{itemize}

\input{table_mwl.tex}

\subsubsection{MWL campaign observations in EHT era}\label{sssec:mwl-eht}

Given that the structure of jets varies both temporally and spatially, it is essential for MWL instruments to measure radiation fluxes from the same celestial direction as simultaneously as possible. 

\begin{figure}[ttt]
    \centering
    \includegraphics[width=\columnwidth]{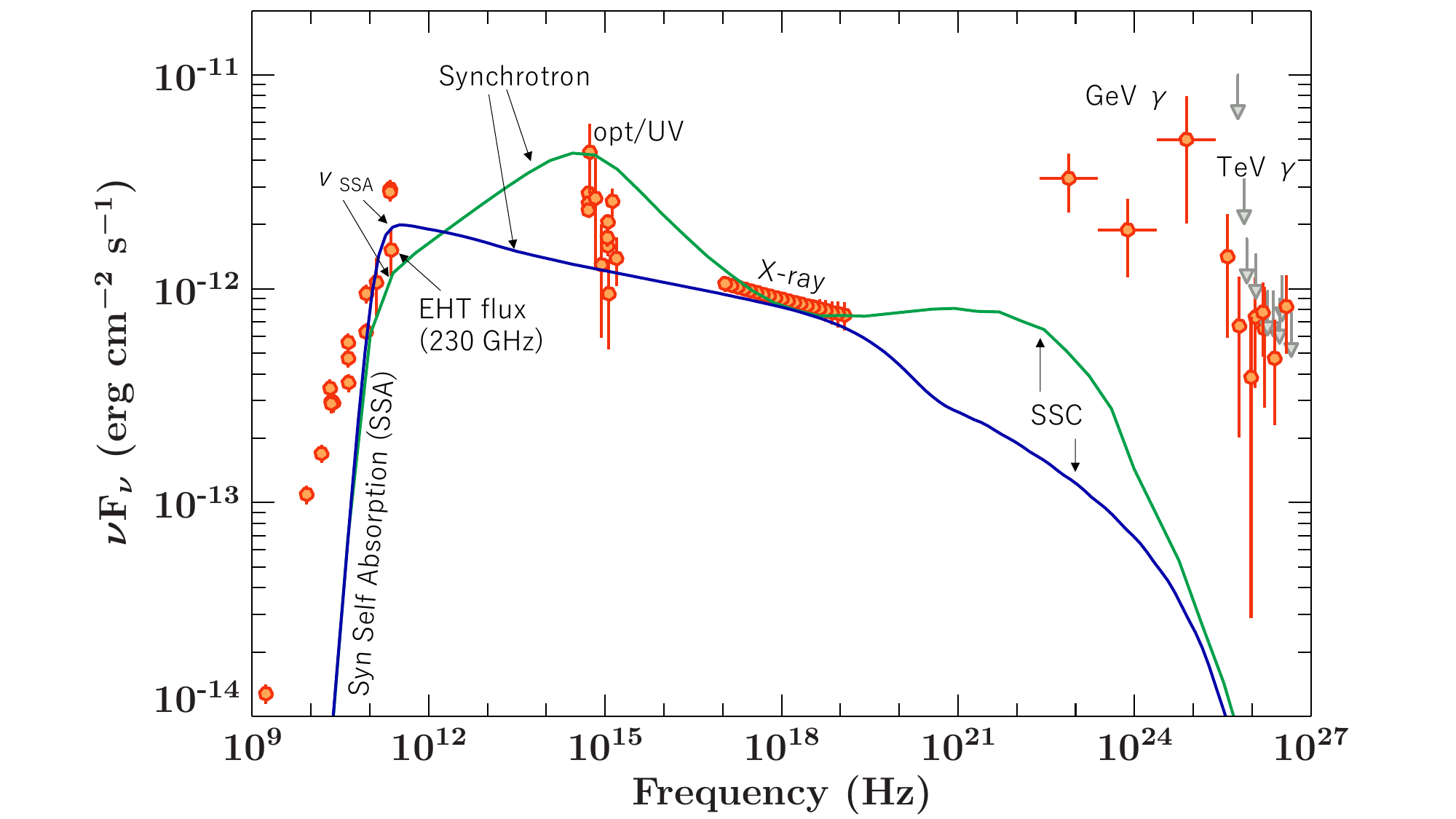}
    \caption{SED fittings of single-zone emission region with the EHT's ring-emission size. 
    Blue and green curves show the SEDs models with different non-thermal electron distribution functions \citep[see details in][]{EHTMWL2021}. In $\gamma$-ray energy bands, one can see the Synchrotron Self-Compton (SSC) emission components although they significantly underestimate the observed $\gamma$-ray flux density. }
    \label{fig:m87SEDfit}. 
\end{figure}

In the 2017 \M87 MWL observations where EHT joined for the first time~\citep{EHTMWL2021}, there were two advancements compared to the prior MWL campaigns: first, quasi-simultaneous observations across various MWL instruments were well organized, minimizing any uncertainty caused by non-simultaneous dataset.
The 2017 M\,87 MWL observing campaign provides the most complete quasi-simultaneous broadband SED of an AGN to date.
Second, the size of the compact (radio) emitting region was tightly constrained thanks to the high-resolution EHT imaging. 
These advantages have allowed us to examine the MWL properties of this active nucleus more accurately. 
As a first step, \citet{EHTMWL2021} modeled the observed SED using simple single zone models with the purpose of highlighting some baseline properties of the peak emission regions (Fig.\ref{fig:m87SEDfit}). The key insights obtained from their SED modeling can be summarized as follows:

\begin{itemize}
    \item When the observed SED is modeled together with the size and mm flux constraints from EHT, the predicted VHE $\gamma$-ray flux from such a compact region significantly underestimates the observed VHE $\gamma$-ray flux. This suggests that the observed VHE $\gamma$-ray radiation is primarily generated in locations different from the EHT region, and challenges the conventional picture that the bulk of observed MWL radiation from radio to $\gamma$-rays originates in a single region.
    \item The synchrotron self-absorption (SSA) turnover frequency $\nu_{\rm SSA}$ of the radio spectrum results in approximately equal to the observing frequency of EHT i.e., 230 GHz. Previous SED modeling studies of M\,87 estimated the magnetic field strength of the emission region to be a few milli Gauss \citep{Abdo09, MAGIC20}. However, the results from \citet{EHTMWL2021} showed a significantly larger value, approximately $B\sim 5$~Gauss, differing by three orders of magnitude.  The difference can be attributed to the difference in the value of $\nu_{\rm SSA}$. Prior to the EHT era, the best-fit $\nu_{\rm SSA}$ in SED models tended to be lower than 230\,GHz since VLBI observations at lower frequencies were still not able to resolve the spatial scales expected from the models. This inevitably led to milli-Gauss order magnetic field strengths.
\end{itemize}

There are a few more recent publications based on TeV $\gamma$-ray data obtained after 2021 \citep{UrenaMena22, HESS24, molero23}. Also, a second EHT-MWL joint campaign was performed in 2018, where M\,87 showed a flaring activity in TeV bands in contrast to the case in 2017 \citep{EHTMWL2024}. These accumulated EHT and MWL data will further refine the localization of VHE $\gamma$-ray emitting region in this source.

The exact physical origin of the \M87 VHE $\gamma$-ray emission remains a topic of debate.
Various potential mechanisms are proposed in the literature, including:
(i) originating from 
magnetospheric vacuum gaps
\citep[e.g.,][]{neronov07, levinson11, Katsoulakos18,kisaka22},
(ii) arising from  magnetic reconnection  processes
\citep[e.g.,][]{ripperda22, hakobyan23}, and 
(iii) stemming from hadronic processes
\citep[e.g.,][]{Reimer04,barkov12, kimura20}.
Further dedicated joint observational efforts between the EHT and VHE observatories will elucidate the physical mechanisms underlying the origin of VHE $\gamma$-ray emission.

\subsection{Magnetic field strengths}\label{ssec:B}

In this section, we review the estimations of some key quantities relevant to the strength of magnetic field based on actual observational data of M\,87.

\subsubsection{Radial profile of magnetic field strength}\label{sssec:Brad}
First, here we overview how the strength of magnetic field ($B$) spatially evolves in the M\,87 jet.
For M\,87, $B$ were estimated at various locations through modelings of diverse observational features such as: spectra of SSA radio core~\citep[e.g.,][]{Reynolds96, Kino2014}; radio spectra of optically-thin jet~\citep{Ro2023a}; brightness temperature of the radio core~\citep{Kim2018}; EHT images~\citep{EHTC2019e,EHTC2021b}; radio-to-$\gamma$-ray broadband SED~\citep[e.g.,][]{georganopoulos05, Abdo09,Prieto16}; core-shift~\citep[e.g.,][]{Zamaninasab14, Jiang21}; variability in radio/X-ray light curves~\citep{harris03,harris09,Acciari09,hada12}.

\begin{figure}[ht]
    \centering
    \includegraphics[width=\columnwidth]{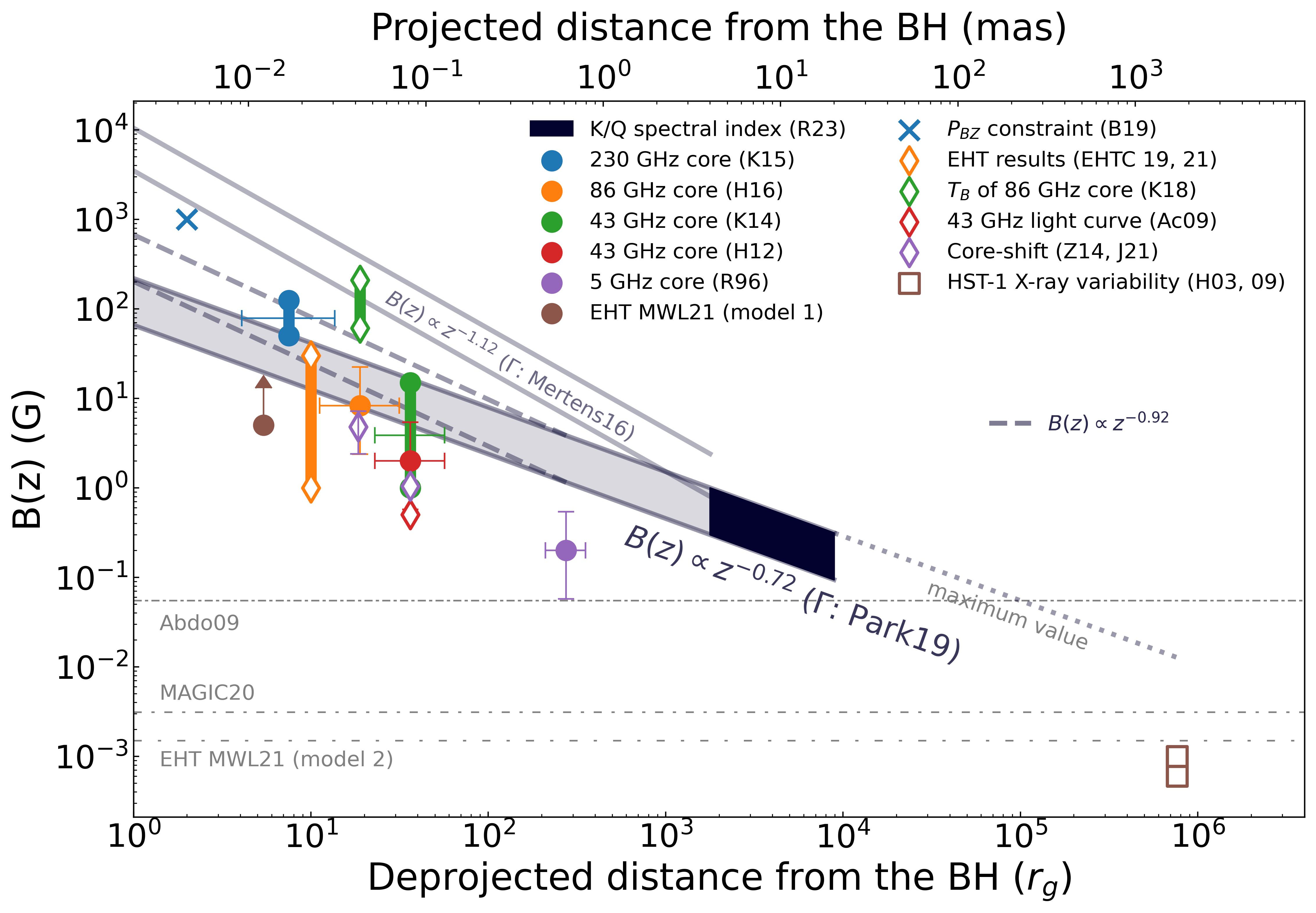}
    \caption{Radial profile of the magnetic field strengths along the M\,87 jet. Here data points are plotted as a function of $r_{\rm g}$ instead of $r_{\rm S}$ presented in the original figure in \citet{Ro2023b}. For more details of descriptions in the figure, see \citet{Ro2023b}. Figure produced by courtesy of Hyunwook Ro.   
    }
    \label{fig:ro2023}
\end{figure}

\citet{Ro2023b} compiled these past $B$ measurements as a function of radial distance from the jet base (Fig.~\ref{fig:ro2023}). Although individual measurements have rather large uncertainties (as large as a factor of $>$10), a collection of independent estimates infer $B$$\sim$1--100\,G levels within a hundred gravitational radii \citep[][]{Kino2015,Hada16,EHTC2021b,EHTMWL2021}. 
When assuming $L_{\rm j} \approx L_{\rm BZ}$ at the jet base of \M87,  the magnetic field strength at the event horizon may exceed $B\gtrsim 10^{2-3}$ G
\citep{blandford19, kino22} (see also Sects.~\ref{label:sssec:phi} and \ref{ssec:Lj}),
\footnote{
\citet{kino22} proposed a slower angular frequency $\Omega_{\rm F}\sim 10^{-2}\Omega_{\rm H}$ than that typically suggested $\Omega_{\rm F}\sim \Omega_{\rm H}/2$ to explain the observed jet velocity profile shown in Fig.~\ref{fig:ACZ}. This slower $\Omega_{\rm F}$ necessitates a larger $B$ at the horizon.
}
although its exact requirement depends on  somewhat ambiguous value of $L_{\rm j}$ (see Sect.~\ref{ssec:Lj}).
On the other hand, $B$ drops to sub-mG levels at a distance of HST-1
\citep{harris03,harris09, giroletti12}. 
It is still uncertain how $B$ evolves between BH to HST-1, but a recent spectral analysis of a pc-scale jet has deduced $B$ and its radial slope at intermediate scales~\citep{Ro2023a}, starting to reveal the global evolution of $B$ in the jet ACZ.

\subsubsection{$U_{\rm e^{\pm}}/U_{B}$ at the jet base}\label{sssec:UeUb}

Next, we discuss another important quantity, $U_{\rm e^{\pm}}/U_{B}$, 
the ratio of the energy density of magnetic field ($U_{B}=B^{2}/8\pi$) to that of nonthermal electron-positron energy density ($U_{\rm e^{\pm}}$) in the jet. 
This ratio at the jet base can be constrained by observed electromagnetic signals, indicating the dynamical significance of the magnetic field in jet launching.
Firstly, we summarize findings from the previous studies based on one-zone models focusing on the opaque SSA sphere. It has long been known that once the angular size and the synchrotron flux from the $\tau_{\rm SSA}\approx 1$ surface are obtained, 
one can uniquely determine $U_{\rm e^{\pm}}/U_{B}$ inside the $\tau_{\rm SSA}\approx 1$ surface \citep[e.g.,][]{kellermann69,blandford78}.
However, spatially resolving jet bases has been challenging due to limited telescope resolution. 
Recent improvements in VLBI observations now enables the application of this method at the base of the M87 jet, suggesting  $U_{\rm e^{\pm}}/U_{B}\ll 1$ \citep{Kino2015}.

Broadband SED fitting of the MWL emission is another
well-known method to investigate $U_{\rm e^{\pm}}/U_{B}$. With the capability to observe AGN jets in the high-energy domain, one can detect not only synchrotron radiation but also inverse Compton scattering components \citep[e.g.,][]{gaidos96,chadwick99,hartman99,abdo10,rieger12,abdo13}
(see Sect.~\ref{ssec:mwl}). 
This allows us to differentiate between  $U_{B}$ and  $U_{\rm e^{\pm}}$, making it possible to determine both of them without degeneracy. From a one-zone fitting to the radio-to-$\gamma$-ray broadband SED, it is suggested that $U_{\rm e^{\pm}}/U_{B}\gg 1$ in the compact emission region of \M87 the jet \citep{Abdo09,MAGIC20}.

Hence, the estimated $U_{\rm e^{\pm}}/U_{B}$ values showed contrasting results even among simple uniform one-zone models.
However, the situation is improving. As explained in Sect.~\ref{ssec:mwl}, the MWL analysis including EHT data seems to gradually mitigate the contradiction \citep{EHTMWL2021}. 
In the compact region near the BH observed by EHT, 
$U_{\rm e^{\pm}}/U_{B}$ tends to decrease and become the order of unity, approaching the situation presumed in GRMHD simulations
(Fig.~\ref{fig:m87SEDfit}). 
On the other hand, the VHE $\gamma$-ray emission region  
is suggested to be $U_{\rm e^{\pm}}/U_{B}\gg 1$, located at a different place away from the BH vicinity.

\subsubsection{$\Phi_{\rm BH}$ at the event horizon}\label{label:sssec:phi}

As summarized in Sects.~\ref{sssec:Brad} and \ref{sssec:UeUb}, 
various observational data colleted over the years for \M87 support the idea of the jet being magnetically driven. Therefore, we now estimate the value of $\Phi_{\rm BH}$, the magnetic flux threading the event horizon of the central BH.
Using Eq.~(\ref{eq:PhiBH}) together with 
the values of $\dot{M}$ (Sect.~\ref{sec:accretion}) and $M_{\rm BH}$ (Sect.~\ref{sec:mass})
obtained for \M87, the magnetic flux threading the event horizon can be
estimated as follows:
\begin{align}
    \Phi_{\rm BH}  
        &\approx 2 \times 10^{33} 
        \left(\frac{\phi_{\rm BH}}{50} \right) 
        \left(\frac{\dot{M}}{10^{-3}~M_{\odot}~{\rm  yr}} \right)^{1/2} \left(\frac{M_{\rm BH}}{6.5\times10^{9} M_{\odot}}\right) \, {\rm G\, cm^{2}} ,
\end{align}
where $\phi_{\rm BH} \approx 50$ is assumed in line 
with recent GRMHD simulations.
This seems to agree  with the sequence of magnetic flux estimated at the downstream of jets 
\citep{Zamaninasab14},
Using the relation $\Phi_{\rm BH} \approx B_{\rm H} r_{\rm H}^{2}$, where $B_{\rm H}$ is the strength of the magnetic field at the event horizon, the magnetic field strength can be estimated as follows:
\begin{align}
    B_{\rm H}  \approx 2 \times 10^{3}
     \left(\frac{\phi_{\rm BH}}{50} \right) 
        \left(\frac{\dot{M}}{10^{-3}~M_{\odot}~{\rm  yr}} \right)^{1/2} \left(\frac{M_{\rm BH}}{6.5\times10^{9} M_{\odot}}\right)
     \, {\rm G}  ,
\end{align}
which can be comparable to the estimates of $B_{\rm H}$ 
mentioned in Sect.~\ref{sssec:Brad}.
Together with the above estimated $\Phi_{\rm BH}$
and  Eq.~(\ref{eq:L_BZ}),
BZ power of M\,87 can be estimated as 
\begin{align}\label{eq:LBZ_M87}
    L_{\rm BZ}  \approx 5 \times 10^{43}
    & \left(\frac{\kappa}{0.05}\right)
        \left(\frac{\Omega_{\rm H}}{9.8\times 10^{-6}~{\rm rad~s^{-1}}}\right)^{2} \nonumber \\
    & \left(\frac{\Phi_{\rm BH}}{2\times 10^{33}~{\rm G~cm^{2}}} \right)^{2}
     \, {\rm erg~s^{-1}}  ,
\end{align}

where $\Omega_{\rm H}$ is given by Eq.~(\ref{eq:Omega_H}) and
here we set $a\approx 0.9 $ and $f(\Omega_{\rm H}) \approx 1$
\citep[e.g.,][]{Tchekhovskoy10}.
In this way, $ \Phi_{\rm BH}$ can establish a connection between the event horizon and the downstream region of the jet.
The estimated $L_{\rm BZ}$ shown in Eq.~(\ref{eq:LBZ_M87}) 
can be comparable to $L_{\rm j}$ from observations
(see Sect.~\ref{ssec:Lj}),
suggesting a capablility to supply the jet power.

\subsection{Estimation of $L_{\rm j}$}\label{ssec:Lj}

As reviewed in Sect.~\ref{ssec:key-lj}, 
it is difficult to put a limit on the total amount of nonthermal electrons (and positrons) because information can only be obtained directly from the jet's electromagnetic observations for a portion of nonthermal electrons (and positrons). Therefore, there have been no significant recent progress in the estimation of $L_{\rm j}$ for M87, and we will rely on the estimates made in relatively past studies.
In the following, we proceed with the review in order of increasing spatial scale.

On several 10 kpc scales,
$L_{\rm j}$ is estimated from the energy required to inflate the observed X-ray cavity and radio lobe structures. 
The previous works suggest  $L_{\rm j}\sim 10^{43-44}~{\rm erg~s^{-1}}$ \citep{owen2000,  young02,allen06,rafferty06}.
The spherical shock model provides an estimate of the
mean power of the shock outburst as $2 \times 10^{43}~{\rm erg~s^{-1}}$ averaged over the past $10^{7}$ year
on $\sim 50$ kpc scale \citep{forman05}.
An application of synchrotron aging model against the Low-Frequency Array (LOFAR) observation of the radio lobe suggests an age of $\sim$40 Myr, which leads to 
$L_{\rm j} \sim (6-10) \times 10^{44}~{\rm erg~s^{-1}}$
\citep{deGasperin12}.

At a distance of about 1\,kpc from the central 
BH, Knot A is known to be a bright optical feature in the jet downstream side. 
\citet{bicknell96} made an estimation of $L_{\rm j}$ using
the Knot A. 
Interpreting knot A as an oblique shock within the jet results in a jet power estimate of
$(1-3) \times 10^{44}~{\rm erg~s^{-1}}$.
The estimated $L_{\rm j}$ is consistent with the above-mentioned jet power required to feed the radio lobe.
In addition, the search for a steady and extended $\gamma$-ray signal around M~87 can constrain the upper limit of cosmic-ray energy density, as discussed in a recent study by the H.E.S.S. collaboration, which addressed the upper limit of $L_{p}$ \citep{HESS23}.

By identifying HST-1 as a recollimation shock, \citet{stawarz06} estimated $L_{\rm j}$
by the pressure balance between
the ambient matter and HST-1.
They derive a jet power of the order of
$10^{44}~{\rm erg~s^{-1}}$.
The estimated $L_{\rm j}$ is again consistent with the above-mentioned jet power required to feed the radio lobe.

In addition, it is worth noting that \citet{broderick15} pointed out that the above-mentioned $L_{\rm j}$ implies a certain minimum $\dot{M}$ and if the central compact object in \M87 were not a BH but had a surface, the emission from the accreting matter onto the surface would exceed the observed fluxes in near-infrared and optical bands, suggesting the presence of an event horizon.


\section{Where do we stand?}\label{sec:wheredowestand}

Through Sect.~\ref{sec:mass}, \ref{sec:accretion} and \ref{sec:jet}, we have reviewed various observational results and progress on the central SMBH, accretion flows and relativistic jets of M\,87. By putting together the accumulated knowledge, here we summarize our current understanding of this active galaxy. We also provide notes on future outlooks, highlighting some remaining questions to be explored in future studies.

\subsection{Key parameters}\label{ssec:m87parameter}

In Table~\ref{tab:param} we summarize estimated values of various parameters and observational quantities for M\,87, including those introduced in Sect.~\ref{sec:key}. The table reflects our current level of understanding about this source. One can see that some of the key quantities are already determined with sufficient accuracy, while some other parameters remain highly uncertain. It is also worth noting that, for some important physical quantities (e.g., the accretion rate, jet shape, jet velocity, magnetic field strength), we are starting to see their spatial evolution over many orders of magnitude in radial distance ($r$) or jet axial distance ($z$), owing to the accumulation of observational data at different scales and the progress of multi-scale studies over the past years. Therefore, we are now beginning to connect the physics between horizon scales and galactic scales, which is essential for a comprehensive understanding of this active galaxy as a whole system. 

\input{table_parameters}

\subsection{What's left, and what are the next steps?}\label{ssec:nextstep}
More than a quarter of a century has passed since the first dedicated workshop on the radio galaxy M\,87 in Tegernsee, Germany in 1997. 
During the conference, some important questions were left unanswered \citep{blandford1999a}: \textit{`Why is M\,87 so dim?'}; \textit{'Does the Jet Power Come from the Disk or the Hole?'}; \textit{'How is the Jet Collimated and of What Is It Made?'}; \textit{'What's Going on in Knot A?'}. We believe that these fundamental quests have been extensively studied both observationally and theoretically since then. In addition, the second M\,87 workshop\footnote{\url{https://events.asiaa.sinica.edu.tw/workshop/20160523/}} in Taipei, Taiwan, in 2016 drove momentum for the imaging of a BH shadow, along with preparations for the 100th anniversary of the discovery of astrophysical jets. Now, we know answers to the questions above and we know that there are SMBHs at the center of galaxies. This is, however, still a good time to be studying M\,87, as Roger Blandford mentioned before.

Below we describe some of the selected remaining challenges or topics that should be explored in future studies of M\,87.  

\begin{itemize}
    \item \textbf{BH spin:} As summarized in Table~\ref{tab:param}, a growing number of studies suggest that the M\,87 SMBH is likely spinning, whereas the exact value of the spin remains highly uncertain. Constraining the spin parameter is crucial for testing the spacetime properties near the SMBH and for testing the jet launching mechanism via BZ processes. To directly constrain the spin, one of the most promising ways would be to accurately measure the shape of a photon ring, where the ring shape is expected to deviate from a pure circular shape as the spin increases (up to $\pm$$\sim$5\%)~\citep{Takahashi2004, Johanssen2010}. The ring-like structure observed by the ground-based EHT baselines at 1.3\,mm is likely still dominated by the emission from the gas surrounding the BH, while the emission from a lensed photon ring becomes significant only at baselines longer than $\sim$20\,G$\lambda$~\citep{Johnson2020J}. To detect signals from such an uncontaminated photon ring, future higher-resolution mm/submm VLBI projects, 
     such as the next-generation EHT (ngEHT\footnote{\url{https://www.ngeht.org/}}), Black Hole Explorer (BHEX\footnote{\url{https://www.blackholeexplorer.org/}}) with a space satellite and/or VLBI at even higher frequencies
     \citep[e.g., GLT;][]{ChenMT23} will be promising.    
    \item BH-jet connection: While the EHT 2017/2018 data clearly detected the BH shadow, they failed to recover the surrounding jet emission mainly due to the limited baseline coverage of the array, leaving the exact connection between the shadow and jet launching still unresolved. Future EHT, ngEHT and/or BHEX observations at 230/345\,GHz with significantly enhanced array capabilities hold promise for detecting both the shadow and the jet-base emission in the same image~\citep[e.g.,][]{EHTCMid2024}, which is also crucial for constraining the BH spin as well as the jet-launching mechanism. This exploration will be further reinforced by complementing multi-frequency high-resolution images at 86\,GHz and 43\,GHz, through which one can trace a full evolution of initial jet formation from $\sim$1\,$r_{\rm S}$ to $\sim$1000\,$r_{\rm S}$.
    \item \textbf{Dynamical/Time-domain view:} Thanks to the accumulation of multi-epoch, multi-wavelength, and multi-scale observational data, now we know that the structure of M\,87 is variable at all spatial scales (from horizon to kpc) and time scales (from day to years). However, none of these observed variable features remain fully understood due to the complex nature of the underlying physics and the general lack of time-domain observational data. Time-domain/dynamical information provides much richer insights into the accretion and jet physics as well as the mechanisms of high-energy emission and particle acceleration processes. Continued multi-epoch, multi-instrument, multi-wavelength observations are highly encouraged to improve our knowledge of the time-domain properties of this object.  
    \item \textbf{Mass transport processes:} 
    Dating back to the mid-1990s, ADAF \citep{narayan1995} was proposed to account for SgrA* and the subsequent development of RIAF with substantial mass loss including the convection \citep{abramowicz2002} and winds \citep{blandford1999} has advanced our understanding of the mass transport towards sub-Eddington BHs including M\,87. A substantial decrease in $\dot{M}$ may be the norm, but we may not have a problem with this fact even if when considering jet production \citep{Ghisellini2005}. When an accreting BH is in the MAD state \citep{Narayan03}, the spinning BH is able to power the jet \citep{blandford77}. However, the angular distribution of the accretion flow and winds remains unclear, so the net amount of mass transport onto the BH has not yet been confirmed at a quantitative level. The issue of mass and energy transport should be examined in multiple dimensions\citep{yuan2014, Yuan2022}. The fate of winds with significant mass loading is of particular interest, and further investigations with Faraday RM measurement are encouraged.
    \item \textbf{Jet structure:} As reviewed in Sect.~\ref{sec:jet}, a growing number of pc-scale VLBI images suggest that the transverse structure of the M\,87 jet is likely more complicated than previously thought, as represented by the spine-sheath-like morphology. Additionally, these relativistic flows are likely surrounded by (invisible) non-relativistic outflows/winds. On kpc scales, high-quality VLA images confirm the double-helix morphology of the large-scale jet \citep[a similar helical structure has recently been suggested also in the pc-scale jet;][]{Nikonov23}.     Revealing the detailed internal structure of the M\,87 jet is essential to better understand the formation, propagation and interaction with the external medium. To facilitate this study, further improvement in both angular resolution and image dynamic range for the extended jet is required. Future deep interferometric imaging observations with the next-generation VLA (ngVLA\footnote{\url{https://ngvla.nrao.edu/}}) and the Square Kilometer Array (SKA\footnote{\url{https://www.skao.int/}}) operated at cm wavelengths will enable us to conduct this kind of study. 
    \item \textbf{Magnetic fields:}    
Clarifying the geometry of magnetic field is crucial for understanding jet formation.
On near-horizon scales,
\citet{chael23} pointed out that the sign of arg($\beta_{2}$) correlates with Poynting flux direction, and the EHT’s measurement of arg($\beta_{2}$)\footnote{The arg($\beta_{2}$) is the second azimuthal Fourier mode of linear polarization in a near-horizon images \citep{palumbo20}.} implies electromagnetic energy outflow.  
\citet{chael23} found that (i) arg($\beta_{2}$) strongly depends on the BH spin, and (ii) arg($\beta_{2}$) shows a radial dependence, evolving rapidly near the inner shadow as fields are wound up. Thus whether this energy outflow is spin-powered or not remains inconclusive.
To test whether the BZ mechanism really is at work or not in the M\,87 jet,  measuring arg($\beta_{2}$) closer to the event horizon is imperative. Higher-dynamic range or/and higher resolution polarization images with ngEHT/BHEX will be essential to clarify the radial dependence of arg($\beta_{2}$).
On a parsec scale view of the M\,87 jet, the linearly polarized emissions are weak and still detected only part of the jet \citep[e.g.,][]{Zavala2002, Park2019a}.
To explore the overall radial evolution of the magnetic field geometry and understand how it evolves from the BH to larger scales, future high-sensitivity VLBI and multi-wavelength polarimetric observations are highly anticipated.

\item \textbf{Particle acceleration and high-energy emission:}  
The origin of energetic particles emitting VHE $\gamma$-rays is of great interest, as described in Sect.~\ref{ssec:mwl}. During the 2018 EHT campaign, flaring episodes constrained the size of the VHE $\gamma$-ray emitting region in SED modeling. However, a one-zone model cannot fully explain the SED, requiring an additional component.
Two processes likely produce the energetic  particles emitting VHE $\gamma$-rays: magnetic reconnection and shock acceleration. 
More sensitive VHE instruments such as the Cherenkov Telescope Array \citep{acharya13,cta19}, may help identify the VHE $\gamma$-ray emitting region due to their better short-timescale sensitivity and spectral resolution.
Investigating EVPA time evolution of the high-energy emission may also help distinguish these scenarios. Magnetic reconnection flares may cause drastic  changes of EVPA, while shock-induced flares may not. 
Future X-ray polarization mission eXTP \citep{Zhang2019} could 
help distinguish these scenarios.

\end{itemize}

\subsection{Broader context: comparison with other sources}\label{ssec:broadcontext}

The extensive M\,87 studies over the past decades have dramatically improved our understanding of various key properties of AGNs/SMBHs. Nevertheless, M\,87 occupies only part of the whole AGN parameter space. To explore the diverse properties of AGNs and how M\,87 fits into the context of a `unified picture' of AGNs, comparison with other sources is essential. 

First, the Galactic Center SgrA* is one of the most frequently compared objects with M\,87 since they are the only two sources where the SMBH shadow is spatially resolved~\citep{EHTC2022a, EHTC2022b, EHTC2022c, EHTC2022d, EHTC2022e, EHTC2022f}. Unlike M\,87, SgrA* does not have a clear jet. This sharp contrast offers a unique opportunity to tackle the question, directly at the horizon scales, of why some SMBHs exhibit powerful jets while others do not. Interestingly, the EHT results of SgrA* favor the MAD state and high spin models, which might be expected to generate a jet efficiently like M\,87. Other primary differences are that SgrA* has a much lower $M_{\rm BH}$ and $\dot{M}$ than M\,87 both in absolute terms and when scaled by mass, which may also limit the jet power. 

Second, while our understanding of weakly accreting SMBHs has greatly improved thanks to the intensive studies on M\,87 (and SgrA*), the physics of accretion/ejection for other types of RLAGNs, especially those with highly accreting SMBHs, remains far less understood. Such sources may produce even more powerful jets, but they are typically much more distant than M\,87. Recent ultra-high-resolution mm-VLBI and space-VLBI observations have started to resolve the inner jet regions within 100-1000\,$r_{\rm S}$ for an increasing number of RLAGN samples~\citep[]{Kim20,Janssen21, Issaoun22, Jorstad23, Fuentes23, Paraschos24}, covering a variety of jet viewing angles, accretion rates, jet powers and nuclear environments. By comparing the observed jet properties of these sources with those of M\,87 at matched gravitational scales, we will be able to better answer the diversity and universality of RLAGN physics across diverse accretion rates. For instance, an ongoing debate is what physics ultimately controls the extension of jet ACZ. As reviewed in Sect.~\ref{sec:jet}, the jet profile of M\,87 is constrained very well and many other AGN jets appear to exhibit similar parabolic-to-conical transitions regardless of AGN types, suggesting that such a jet transition is universal. However, while the M\,87 jet breaks at the Bondi radius, a majority of other jets seem to break at distances different from their Bondi scales~\citep[e.g.,][]{kovalev2020, Boccardi21, okino2022}. Thus, it is worth continuing to explore how magnetically-driven jets evolve under the stratified atmosphere governed by SMBHs in future observations. This would require more samples with a broad range of AGN parameter space. 

\section{Summary}\label{sec:summary}

In this review, we provided a comprehensive overview about the recent observational progress of M\,87 from galactic scales to event horizon scales, obtained across various wavelengths from radio to TeV $\gamma$-rays. Here we conclude our review by itemizing some key points. 

\begin{itemize}
    \item The mass of M87's central SMBH has been estimated using various methods, including stellar dynamics, gas dynamics, and BH shadow measurements. The current plausible value would be $M_{\rm BH}\sim(6-7)\times 10^9M_{\odot}$ that were consistently obtained from the stellar dynamics and the BH shadow image. 
    \item The nature of accretion flows from Bondi scales to horizon scales was reviewed. The highly sub-Eddington accretion rate at Bondi scales makes BH accretion of M\,87 RIAF-type flows. Putting all $\dot{M}$ estimates at different scales together, substantial decrease in $\dot{M}$ towards smaller radii is indicated. This suggests that only a fraction of accretion energy available at large scales is actually supplied to the central SMBH, and significant mass loss processes are taking place between SMBH and Bondi scales. The observed Faraday RM profiles at parsec scales suggest substantial winds from hot accretion flows, which may be responsible for the reduction of mass supply at small radii. 
    \item We also highlighted the complexity and diversity of phenomena associated with the relativistic jet in M\,87. The collimation profile of this jet is now quite well established over 10$^7$ orders of magnitude in distance. Although still relatively less defined compared to the jet shape, our observational understanding of the jet dynamics (both along and perpendicular the jet axis) has also been significantly updated in recent years thanks to various intensive VLBI monitoring programs. Furthermore, recent high-dynamic-range radio imaging both at pc and kpc scales has revealed the transverse or internal jet structures (e.g., spine-sheath and helical structures) in great detail, highlighting the complexity of jet physics. 
    \item An increasing number of independent studies observed at different spatial scales and wavelengths consistently suggest that the M\,87 central SMBH is spinning. The spinning BH likely plays a key role in generating the observed relativistic jet of M\,87, through the Blandford \& Znajek processes. Nevertheless, the exact value of the spin remains less constrained and stands as a crucial question to be explored in future studies.   
    \item Polarization structures of M\,87 are now revealed in great detail at all scales from horizon to kpc scales, providing additional insights about the physical properties of accretion flows and relativistic jet, as well as the structure of magnetic fields associated with them. The observed polarization characteristics associated with the EHT ring strongly favor MAD states, suggesting the presence of dynamically important magnetic fields near BH, which can be also responsible for producing the observed powerful jet. The organized MWL polarization properties observed in the large-scale jet indicate that magnetic fields continue to play a critical role in determining the structure of the M\,87 jet even at kpc scales.   
   
    \item We also summarized the broadband MWL observing campaigns conducted in the past. With the exception of the 2005 event where HST-1 clearly exhibited MWL ourbursts, most of high-energy (X-ray and $\gamma$-ray) events seem to be associated with the nucleus rather than far from the BH. Nevertheless, the exact location and mechanisms of the $\gamma$-ray emission are still a matter of debate. Continued efforts of coordinated simultaneous MWL observations along with high-resolution radio instruments are highly encouraged.  

    By combining the numerous pieces of knowledge summarized above, we are beginning to seamlessly trace a progression of key physical processes that characterize an AGN, from accretion onto the SMBH to jet launching and propagation beyond the host galaxy, along with energy dissipation through intense radiation. The study of M\,87 has demonstrated that these processes are tightly interrelated across a wide range of physical scales, underlining the importance of multi-scale, multi-faceted approaches to fully disentangle this intricate system.

    \item Despite the tremendous advances in our knowledge of this object, several outstanding questions remain. Next-generation instruments planned for the next decade will be crucial in addressing these remaining questions. It should also be emphasized that comparing M\,87 with other sources at matched gravitational scales is essential for understanding the diverse properties of AGNs. 
\end{itemize}

\bigskip
\bigskip

\backmatter

\bmhead{Acknowledgments} We thank the anonymous referee and the journal editors for their valuable comments and suggestions, which significantly improved the manuscript. We thank Juan Carlos Algaba, Alice Pasetto and Hyunwook Ro for producing Fig.\ref{fig:m87mwl}, \ref{fig:kpc} and \ref{fig:ro2023} for this review, respectively. We thank Alexei Nikonov for providing the data that allowed us to create Fig.\ref{fig:cj} for this review. We also thank Derek Ward-Thompson, Kazi Rygl, Bong Won Sohn, and Victor Barbosa Martins for their reading and commenting on the manuscript. Part of this work was supported by the MEXT/JSPS KAKENHI (grants 21H04488, 21H01137, 22H00157 and 24K07100), the Mitsubishi Foundation grant (grants 201911019 and 202310034), the National Science and Technology Council grant (NSTC 111-2112-M-001-041, NSTC 111-2124-M-001-005, and NSTC 112-2124-M-001-014) and the Academia Sinica Career Development Award grant AS-CDA 110-M05.

\clearpage
\phantomsection
\addcontentsline{toc}{section}{References}
\bibliography{ref_m87}


\end{document}

%% file: table_symbols.tex
\begin{longtable}{ll}
  \caption{Symbols and abbreviations used in this review \label{tab:symbols}} \\
   \hline   
   \multicolumn{2}{l}{Quantities} \\ 
   \endfirsthead
    \multicolumn{2}{l}{\small\it Continued}\\
    \hline \\
    \endhead
    \hline \\
    \endfoot
    \hline \\
    \endlastfoot
       $D$ & Distance to a given source \\
       $M_{\odot}$ & Solar mass \\
       $M_{\rm BH}$ & Black hole mass \\
       $a$    & Dimensionless spin parameter\\
       $J$    & Black hole angular momentum\\
       $r_{\rm g}$    & Gravitational radius\\
       $r_{\rm H}$    & Radius of the event horizon\\
       $r_{\rm E}$    & Radius of the ergosphere \\
       $r_{\rm S}$    & Schwarzschild radius\\
       $R_{\rm tr}$    & Transition radius from an ADAF to a standard accretion disk\\
       $\Omega_{\rm H}$    & Angular frequency of the event horizon\\
       $\Omega_{\rm F}$    & Angular frequency of magnetic field line\\
       $r_{\rm B}$    &  Bondi accretion radius \\
       $r_{\rm c}$   & Core radius of ambient gas\\
       $c_{\rm s}$    & Adiabatic sound speed \\
       $T_{\rm i}$ (ion)    & Ion temperature\\
       $T_{\rm e}$ (electron)    & Electron temperature\\
       $\dot{M}$    & Mass accretion rate\\
       $\dot{M_{\rm B}}$    & Bondi accretion rate\\
       $n_{\rm e}$    & Electron number density \\
       $\rho$    &  Mass density  \\
       $\dot{M}_{\rm Edd}$    & Eddington mass accretion rate\\
       $\dot{m}$    & Mass accretion rate normalized by $\dot{M}_{\rm Edd}$\\
       $\dot{m}_{\rm crit}$    & Critical value (upper limit) of $\dot{m}$ \\
       $\alpha$    &  Viscosity parameter in an accretion flow\\
       $\epsilon_{\rm acc}$    & Accretion efficiency \\
       $\Phi_{\rm BH}$    & Magnetic flux threading the event horizon \\
       $\phi_{\rm BH}$    & Dimensionless magnetic flux threading the event horizon  \\
       $L_{\rm j}$    & Total power of the jet\\
       $L_{\rm rad}$    & Radiation power of the jet \\
       $L_{\rm p}$    & Proton power of the jet\\
       $L_{\rm e^{\pm}}$    & Electron-positron pair power of the jet\\
       $L_{\rm Poy}$    & Poynting power of the jet\\
       $L_{\rm BZ}$    & Power supplied by Blandford \& Znajek process\\
       $\theta_{\rm j}$    & Jet opening angle \\
       $\theta_{\rm view}$     & Jet viewing angle \\
       $\Gamma$     & Lorentz factor \\
       $\beta$     & Jet speed divided by the speed of light \\
       $\sigma$   & Magnetization parameter  \\
       $\delta$    & Doppler factor of the emission region\\
       $\nu_{\rm SSA}$    & Turnover frequency caused by SSA \\
       $U_{\rm e^{\pm}}$    & Nonthermal electron-positron energy density\\
       $U_{B}$    & Energy density of magnetic field\\
       $\tau_{\rm SSA}$    & Optical depth for synchrotron self-absorption\\
                           &  \\
   \multicolumn{2}{l}{Acronyms for objects, physical quantities and states} \\ 
       ACZ & Acceleration and collimation zone \\       
       ADAF & Advection-dominated accretion flow \\       
       ADIOS &  Adiabatic inflow-outflow solution\\       
       AGN & Active galactic nucleus \\
       BLO & BL\,Lac object \\       
       BZ  & Blandford \& Znajek  \\ 
       CD & Current-driven \\
       HD &  Hydrodynamics\\       
       EVPA & Electric vector polarization angle \\       
       FR &  Fanaroff-Riley\\ 
       FR-I & Fanaroff-Riley Class I \\
       FR-II & Fanaroff-Riley Class II \\
       FSRQ & Flat-spectrum radio quasar \\       
       GR & General relativity \\       
       GRRT & General relativistic radiative transfer \\       
       GRMHD & General relativistic magnetohydrodynamics \\       
       IGM & Intergalactic medium \\       
       ISCO & Innermost stable circular orbit \\       
       ISM & Interstellar medium \\       
       JCB & Jet collimation break \\       
       KH &  Kelvin-Helmholtz\\       
       MAD &  Magnetically-arrested disk \\       
       MFPA & Magnetic field position angle \\
       MHD &  Magnetohydrodynamics \\       
       MWL & Multi-wavelength \\       
       NLSy1 & Narrow-line Seyfert 1 \\       
       RIAF & Radiatively inefficient accretion flow \\       
       RM & Rotation measure \\       
       RQ & Radio-quiet \\
       RL & Radio-loud \\
       SANE & Standard accretion and normal evolution \\       
       SED & Spectral energy distribution \\       
       SMBH & Supermassive black hole \\
       SSA & Synchrotron self-absorption \\
       UV & Ultraviolet \\    
       VHE & Very-high-energy \\       
           &  \\
       \multicolumn{2}{l}{Acronyms for observing facilities, telescopes and instruments}  \\      
       ALMA & Atacama Large Millimeter/submillimeter Array  \\       
       BHEX & Black Hole Explorer \\
       CFHT & Canada-France-Hawaii Telescope \\       
       EAVN & East Asian VLBI Network \\       
       EHT & Event Horizon Telescope \\       
       EVN & European VLBI Network \\       
       eXTP & enhanced X-ray Timing and Polarization Mission \\       
       GLT &  Greenland Telescope\\       
       GMVA & Global Millimeter VLBI Array \\       
       HAWC & High-Altitude Water Cherenkov Observatory\\       
       HST & Hubble Space Telescope \\       
       HEGRA & High-Energy-Gamma-Ray Astronomy  \\       
       H.E.S.S & High Energy Stereoscopic System \\ 
       LOFAR & Low Frequency Array \\
       ngEHT & next-generation Event Horizon Telescope \\
       ngVLA & next-generation Very Large Array \\
       MAGIC & Major Atmospheric Gamma-ray Imaging Cherenkov Telescope \\       
       SKA & Square Kilometer Array \\       
       SMA & Submillimeter Array \\       
       STIS & Space Telescope Imaging Spectrograph \\       
       VLA & Very Large Array \\
       VLBA & Very Long Baseline Array \\
       VLBI & Very long baesline interferometry \\       
       VERA & VLBI Exploration of Radio Astrometry \\       
       VERITAS & Very Energetic Radiation Imaging  Telescope Array System \\  
       VLT & Very Large Telescope \\       
       VSOP & VLBI Space Observatory Programee
\end{longtable}

%% file: table_mwl.tex
\begin{table}
  \begin{center}
  \caption{Summary of M\,87 quasi-simultaneous MWL observations \label{table:VHEflare}}
    \begin{tabular*}{\textwidth}{@{\extracolsep{\fill}}ccccc}\hline
        Time & Radio/VLBI & X-ray &VHE& {References}\\
      \hline
        2005       & HST-1   & HST-1   & flare & b,c \\
        2008 Feb   & nucleus & nucleus & flare & d \\
        2010 Apr   & nucleus & quiet   & flare & e \\
        2012 Mar   & nucleus & quiet   & elevated flux$^{a}$ & a,f,g,h \\
        2017 Apr   &  quiet  & quiet   & quiet & i \\
       2018 Apr   &  quiet  & nucleus   & flare & j \\
       \hline
    \end{tabular*}
    \begin{tablenotes}
    \item (a): A recent report indicates a quiescent state \citep{MAGIC20}; (b) \citet{Cheung07}; (c) \citet{harris09}; (d) \citet{Acciari09}; (e) \citet{Abramowski12}; (f) \citet{Beilicke12}; (g) \citet{hada14}; 
    (h)\citet{akiyama15};
    (i) \citet{EHTMWL2021} and references therein; (j) \citet{EHTMWL2024} and references therein. 
    \end{tablenotes}
  \end{center}
\end{table}

%% file: table_parameters.tex
\begin{table}[htbp]
  \begin{center}
  \caption{M\,87 parameters \label{tab:param}}
    \begin{tabular*}{\textwidth}{lll}\hline
       Parameters/quantities & Reference values & Related sections/figures \\ \hline 
       \multicolumn{3}{l}{\bf Key parameters introduced in Sect.~\ref{sec:key}} \\  
       $M_{\rm BH}$ & $(6-7)\times10^9\,M_{\odot}$ & Sect.~\ref{ssec:mass} \\
       $a$ & likely $\ne 0$ &  Sect.~\ref{ssec:spin}, \ref{sssec:testmagacz}, \ref{sssec:oscillation}, \ref{ssec:spine-sheath}\\
       $\dot{M}$ (near horizon) & $(3-20)\times 10^{-4}\,M_{\odot} {\rm yr}^{-1}$ & Sect.~\ref{ssec:accretion_horizon}\\
       $\dot{M}$ (at Bondi radius) & $\sim$0.2\,$M_{\odot}{\rm yr}^{-1}$ & Sect.~\ref{ssec:accretion_bondi}\\
       $\Phi_{\rm BH}$ & $\sim$$10^{33}~{\rm G~cm^{2}}$ & Sect.~\ref{ssec:B}\\
       $L_{\rm j}$ & $\sim$$10^{(43-45)}~{\rm erg~s^{-1}}$ & Sect.~\ref{ssec:Lj}\\
           &  &\\
       \multicolumn{3}{l}{\bf Other observational parameters/quantities} \\
       Source distance $D$  & $\sim$16.8\,Mpc & Sect.~\ref{ssec:m87}, \ref{ssec:mass}  \\
       Bondi radius $r_{\rm B}$  & $\sim$220\,pc ($\sim$$3.5\times 10^5\,r_{\rm s}$)  & Sect.~\ref{ssec:accretion_bondi}  \\ 
       X-ray core luminosity   & $\sim$$7\times 10^{40}$\,erg\,s$^{-1}$ & Sect.~\ref{ssec:accretion_bondi} \\
       Core ring diameter (1.3\,mm)  & $\sim$42\,$\mu$as & Sect.~\ref{ssec:mass}, \ref{ssec:accretion_horizon} \\
       Core ring diameter (3.5\,mm)  & $\sim$64\,$\mu$as & Sect.~\ref{ssec:accretion_horizon} \\
       Jet viewing angle & $\sim$14--20$^{\circ}$  & Sect.~\ref{sssec:acceleration}, \ref{ssec:cj}, \ref{ssec:hst-1}\\
       Jet position angle & $\sim$284--292$^{\circ}$ (nominal 288$^{\circ}$)  & Sect.~\ref{sssec:oscillation}, \ref{ssec:kpc} \\
       Jet radius ($z < r_{\rm B}$) & $\propto z^{0.58}$  & Sect.~\ref{ssec:collimation}, Fig.~\ref{fig:ACZ} \\
       Jet radius ($z > r_{\rm B}$) & $\propto z^{1.04}$  & Sect.~\ref{ssec:collimation}, Fig.~\ref{fig:ACZ} \\
       Core-shift (1-88\,GHz) &    $\propto \nu^{-(0.9-1.0)}$  &  Sect.~\ref{ssec:core}      \\
       Maximum apparent jet speed & $\sim$6\,$c$  & Sect.~\ref{sssec:acceleration}, \ref{ssec:hst-1}, \ref{sssec:kpc-1}\\
       Maximum $\lvert {\rm RM} \lvert$  & a few $\times 10^{5}\,{\rm rad\,m^{-2}}$ & Sect.~\ref{ssec:accretion_withinbondi}, \ref{ssec:polarization}  \\
           &  &  \\
       \multicolumn{3}{l}{\bf Radial ($r$) / jet-axial ($z$) evolution*} \\
       $\dot{M}(r)$ &   & Sect.~\ref{sec:accretion}, Fig.\,\ref{fig:Mdot_sum}     \\
       ${\rm RM}(r)$ &      &  Sect.~\ref{ssec:accretion_withinbondi}, \ref{ssec:polarization}    \\
       $\rho (r)$  &     &  Sect.~\ref{ssec:accretion_withinbondi}      \\ 
       $\Gamma\beta(z)$    &      & Sect.~\ref{sssec:acceleration}, \ref{sssec:kpc-1}, Fig.\,\ref{fig:ACZ}       \\
       $B(z)$    &     &   Sect.~\ref{ssec:B}, Fig.\,\ref{fig:ro2023}      \\ \hline
      \end{tabular*}
      \begin{tablenotes}
        *: Here we do not put reference values since these quantities significantly evolve with distance. Instead, the readers may refer to related sections/figures and references therein for details.  
      \end{tablenotes}
  \end{center}
\end{table}